\documentclass[twocolumn]{aastex63}
\usepackage{amsmath}
\usepackage{url}

\newcommand{\ergs}{erg s$^{-1}$}
\newcommand{\kms}{km s$^{-1}$}

\received{January 01, 2020}
\revised{January 10, 2020}
\accepted{\today}
\submitjournal{ApJ}

\shorttitle{GMRT observations of GRB 171205A}
\shortauthors{Maity \& Chandra}

\begin{document}

\title{1000 days of lowest frequency emission from the low-luminosity GRB 171205A}

\author[0000-0002-4682-6970]{Barun Maity}
\affiliation{National Centre for Radio Astrophysics, TIFR, Pune University Campus, Post Bag 3, Pune 411 007, India}
\email{bmaity@ncra.tifr.res.in}
\author[0000-0002-0844-6563]{Poonam Chandra}
\affiliation{National Centre for Radio Astrophysics, TIFR, Pune University Campus, Post Bag 3, Pune 411 007, India}
\email{poonam@ncra.tifr.res.in}

\begin{abstract}

 We report the lowest frequency measurements of  gamma-ray burst (GRB) 171205A with the upgraded Giant Metrewave Radio Telescope (uGMRT) covering a frequency range from 250--1450 MHz and  a period of $4-937$ days.
It is the first GRB afterglow detected at 250--500 MHz frequency range and  the second brightest GRB detected with the uGMRT. 
 Even though the GRB is observed for nearly 1000 days, there is no evidence of  transition to non-relativistic regime. 
We also analyse the archival {\it Chandra} X-ray data on day $\sim 70$ and day $\sim 200$.  We also find no evidence of a jet break
from the analysis of combined data. We  fit  synchrotron afterglow emission arising from a relativistic, isotropic, self-similar deceleration as well as from a shock-breakout of 
wide-angle cocoon. 
Our data also allow us  to discern the nature and the density of the circumburst medium. We find that the density profile deviates from a standard constant density medium and suggests that the GRB exploded in a stratified wind like medium.
Our analysis shows that the lowest frequency measurements covering the absorbed part of the  light curves are critical to unravel the GRB environment.
Our data combined with other published measurements indicate that the radio afterglow has contribution from two components: a weak, possibly slightly off-axis
jet and a surrounding wider cocoon, consistent with the results of  \cite{izzo+19}. The cocoon emission likely dominates at early epochs, whereas the  jet starts to dominate at later epochs, resulting in flatter
radio lightcurves.

\end{abstract}

\keywords{Gamma Ray Bursts---Afterglow---Synchrotron emission}

\section{\textbf{Introduction}} \label{sec:intro}
Gamma Ray Bursts (GRBs) are the most energetic flashes of gamma rays,
with  $T_{90}$ duration (time interval between which 5\% to 95\% of fluence  is collected by the  detector)  ranging between a few milli-seconds to thousands of seconds  \citep{GRB}.  The GRBs can be classified
in two broad classes i.e short/hard GRBs with duration less than 2 secs, and long/soft GRBs of duration greater than 2 secs \citep{1993ApJ...413L.101K}.  According to well accepted theories, most of the long soft GRBs originate from gravitational collapse of massive stars 
(collapsar model);  and short hard GRBs result from explosive binary compact object mergers \citep{woos}.  
The GRBs from both channels of formation power relativistic collimated jets which give Doppler-boosted high luminosities in gamma-rays. GRBs are cosmological events with an average redshift $z\approx2.2$ \citep{fynbo+09}, and have isotropic-equivalent gamma ray luminosities ($L_{\rm iso}$) of the order of $10^{51}$ to $10^{53}$ erg s$^{-1}$.

However, a handful of long/soft GRBs with spectroscopically identified supernovae have been discovered with luminosities that are 
3--5 orders of magnitude lower than the average, i.e.   \citep[$L_{\rm iso} \le 10^{48.5}$ \ergs;][]{schulze+14, cano+17}. 
Their low luminosities allow them to be detected only at low redshifts, though they may be 10--100 times more abundant than regular GRBs
\citep{schmidt+01}. 
 The 
prompt light-curves of typical 
low-luminosity GRBs are smooth and  spectra have a single peak
with the peak energy generally below $\sim$50 keV, which softens further with time  \citep{ns12, cano+17}. 
The radio afterglow of these GRBs tend to indicate similar energy content  in  mildly relativistic ejecta \citep{kulkarni}. 
Many of these GRBs are associated with broad-line Type Ic supernovae. 
 The list of such  GRBs-supernovae include some of the well studied cases such as, GRB
980425/SN 1998bw \citep[$z=0.00866$,][]{galama+98}, GRB 030329/SN 2003dh \citep[$ z = 0.1685$,][]{hjorth+03}, 
GRB 031203/SN 2003lw \citep[$z=0.1055$,][]{malesani+04},  GRB 060218/SN 2006aj \citep[$z=0.0335$,][]{campana+06}, GRB 100316D/SN
2010bh \citep[$z=0.0591$,][]{starling+11}, GRB 111209A/SN 2011kl \citep[$z=0.677$,][]{gao+16}, GRB 120422A/SN 2012bz \citep[$z = 0.283$,][]{melandri+12},
GRB 130427A/SN 2013cq \citep[$z = 0.3399$,][]{melandri+14}, GRB 130702A/SN 2013dx \citep[$z=0.145$,][]{cenko+13}, GRB 161219B/SN  2016jca \citep[$z=0.1475$,][]{cano+17b}, GRB 171010A/SN 2017htp \citep[$z=0.33$,][]{melandri+19},  GRB 190829A/ SN 2019oyu \citep[$z=0.08$,][]{terreran+19}, although only the five are nearby, $z\lesssim0.1$ \citep[][and references therein]{cano+17}.

A significant amount of work has gone  towards understanding 
whether the  low-luminosity GRBs are simply the low-energy counterparts of the cosmological GRBs, or have a different 
emission mechanism. Many low-luminosity GRBs do not follow the $E_{\rm iso}-E_p$ Amati relation \citep{amati+02}, indicating  that their
emission mechanism should be different from that of canonical, more distant  GRBs. Technically an off-axis jet can also explain low-luminosity emission from GRB, but predicts an achromatic  steepening of the light curve, absent in many 
low-luminosity GRBs.
There have been suggestions that in contrast to the emission from an ultra-relativistic jet driven by a  central engine, these low-luminosity GRBs are
powered by  shock breakouts \citep{kulkarni, ns12, nakar15, bd+15, suzuki+17}. In some cases  observations have suggested a mildly relativistic blast wave
being responsible for producing the  radio afterglow. This has supported the relativistic shock breakout model, e.g. GRB 980425 \citep{kulkarni}.
The shock-break out model got further support in  case of GRB 060218, in which a thermal component was also seen which cooled and shifted to optical/UV band with time. This was interpreted to be arising from the break out of a shock driven by a mildly relativistic supernova shell in the progenitor wind \citep{campana+06}, although  late time photospheric emission from a  jet
\citep{fw13}, or thermal emission from  a cocoon \citep{ss13} can also explain it.
  \citet{bromberg+11} have investigated whether the low-luminosity GRBs launch relativistic jets like their high energy counterparts, but incur
resistance by the stellar envelopes surrounding their progenitor stars.  They found that some low-luminosity GRBs have much shorter durations
 compared to the jet breakout time, This is inconsistent with the collapsar model which is largely successful in explaining the cosmological GRBs \citep{GRB}.

While \citep{bd+15} and \citep{ns12} have developed spherical relativistic shock break out models in 
context of low-luminosity GRBs,
\citep{nakar15} 
addressed some of the problems of spherical shock-break out model by introducing a low mass optically thick stellar envelope surrounding the progenitor star.
 In this
model, the explosion powering the low-luminosity GRBs was not the spherical breakout of the supernova shock, but by a
 jet that gets choked in the envelope and powers quasi-spherical explosion.
To extend this idea further,  \citep{np17} considered a cocoon breakout model.
In this model, as the GRB jet pushes through the stellar material, it heats the surrounding gas and produces a high-pressure sub-relativistic
cocoon, which at the time of breakout, produces a relatively faint flare of $\gamma$-rays.
This break out  will not be as spherical as a supernova break-out, but will be wider than a jet. In this model, 
interaction of the cocoon with the surrounding medium can give rise to a late time radio and X-ray afterglow.
However, \citep{ic16} have provided an alternative mechanism, where the composite emission of GRB 060218/SN 2006aj could be explained by a weak jet,
along with a quasi-spherical supernova ejecta.

 llGBRs also have a radio afterglow, which indicates a comparable energy in mildly relativistic ejecta (Kulkarni et al. 1998; Soderberg et al. 2004, 2006; Margutti et al. 2013).
 The lack of bright, late-time, radio emission from ll-GRBs strongly constrain the total energy of any relativistic outflow involved in these events 
 (Waxman 2004; Soderberg et al. 2004, 2006b). Additionally, statistical arguments rule out the possibility that ll-GRBs are regular LGRBs viewed at a 
 large angle (e.g., Daigne \& Mochkovitch 2007). Thus, if ll-GRBs are generated by relativistic jets these jets must be weak and have a large opening 
 angle. The bursts with $L_\gamma > 2 \times10^{48}$ \ergs\ are considered as regular GRBs and are separated into LGRBs and SGRBs according to the 
 standard criterion of whether T90 in the observer frame is above or below 2 s (Bromberg, Nakar, Piran 2011).

GRB 171205A is a nearby low-luminosity GRB with a $T_{90}$ duration of almost 189.4 secs \citep{delia+18}. It was first discovered by the  Burst Alert Telescope (BAT) onboard {\it Swift} on 5th December, 2017  \citep{discovery}. It has a redshift of 0.0368 \citep{discovery,redshift}.  The isotropic energy release in gamma ray band at GRB rest frame was $2.18_{-0.50}^{+0.63}\times 10^{49}$ ergs \citep{delia+18}. The host of this event was a bright spiral galaxy named 2MASX J11093966-1235116 \citep{redshift}, with a mass of the order of $10^{10}M_{\odot}$ and a star formation rate of $3\pm1M_{\odot}/\rm{yr}$ \citep{perley}. An emergent supernova event (SN 2017iuk) was seen three days after the burst \citep{delia+18}. 

There are a handful of observations for GRB 171205A  from mm to radio bands. It was detected by  \citet{postigo}  using  Northern Extended Millimeter Array (NOEMA) with a flux density of $\sim 35$ mJy at 150 GHz after 20.2 hours of the burst.
 Atacama Large Millimeter/Submillimeter Array (ALMA) detected a bright afterglow of significance more than 100$\sigma$ at 92 GHz and 340 GHz on 10-11 December, 2017 \citep{alma}. 
RATAN-600 radio telescope detected it at 4.7 and 8.2 GHz bands  during 9th December to 16th December, 2017 \citep{RATAN}. Karl G. Jansky Very Large Array (VLA)  also observed the afterglow in the  frequency range 4.5 to 16.5 GHz  \citep{vla}. This GRB also had a Very Long Baseline Array (VLBA) detection at different frequencies \citep{vlba}. 
These  observations  showed a steeply rising spectrum($\propto\nu^{2}$) at low frequency which indicates a synchrotron self-absorbed spectrum. 
 This is the first GRB for which \citet{urata} claimed to have detected  polarization at the ALMA 90GHz frequency, though this claim has been disputed by \citet{laskar+20}. The upgraded Giant Metrewave Radio Telescope (uGMRT) first detected it on 20th December, 2017 at 1400 MHz \citep{gmrt} after a non detection  on 10th and 11th December, 2017 \citep{gmrt1}. The observed flux density at that epoch was $782\pm57 \rm{\mu Jy}$.

Radio afterglow emission from GRBs evolve slowly which gives us the opportunity to observe it for a long time and obtain  the distribution of the kinetic
energy in the velocity space. Since this distribution is different for various models, especially central engine driven versus shock break out, 
radio observations provide unique opportunity to  distinguish between various
emission models \citep[e.g. ][]{kulkarni}. 
Additionally, the early radio emission is likely to be absorbed via synchrotron self-absorption (SSA), and thus constrain the circumburst medium (CBM) density \citep{Chandra}. 
The late time radio observations in the Newtonian limit, when the jet becomes sub-relativistic, are nearly independent of jet geometry and measure the kinetic energy of the afterglow accurately \citep{frail}.

In this paper we present low frequency observations of GRB 171205A taken with the uGMRT for around 1000 days. We summarize our observations and  data analysis in \S 
\ref{sec:style}. We discuss our model in \S \ref{sec:model} and results in \S \ref{sec:results}. In \S \ref{sec:discussion}, we discuss the properties of 
GRB 171205A in conjunction with published measurements at higher frequencies and present our main conclusions.
Unless otherwise stated, we assume a cosmology with $H_0=67.3$ km\,s$^{-1}$\,Mpc$^{-1}$, 
$\Omega_m=0.315$, $\Omega_\Lambda=0.685$ \citep{planck14}.

\section{\textbf{Observation and Data Analysis}} \label{sec:style}

The GMRT observed GRB 171205A starting 2017 December 10 and continued observing until 2020 June 26. The observations 
were taken in band 5 (1000--1450\,MHz), band 4 (550--900\,MHz) and band 3 (250--500\,MHz).
The bandwidth for band 4 and band 5 was 400 MHz while for band 3 it was 200 MHz. The duration of each observation was around 2-3 hours including overheads  (on source time 1.5 hours). We observed flux density calibrators 3C286 and  3C48; and a phase calibrator J1130-148. Flux calibrators were also used as bandpass calibrators.

\begin{figure*}
\gridline{\fig{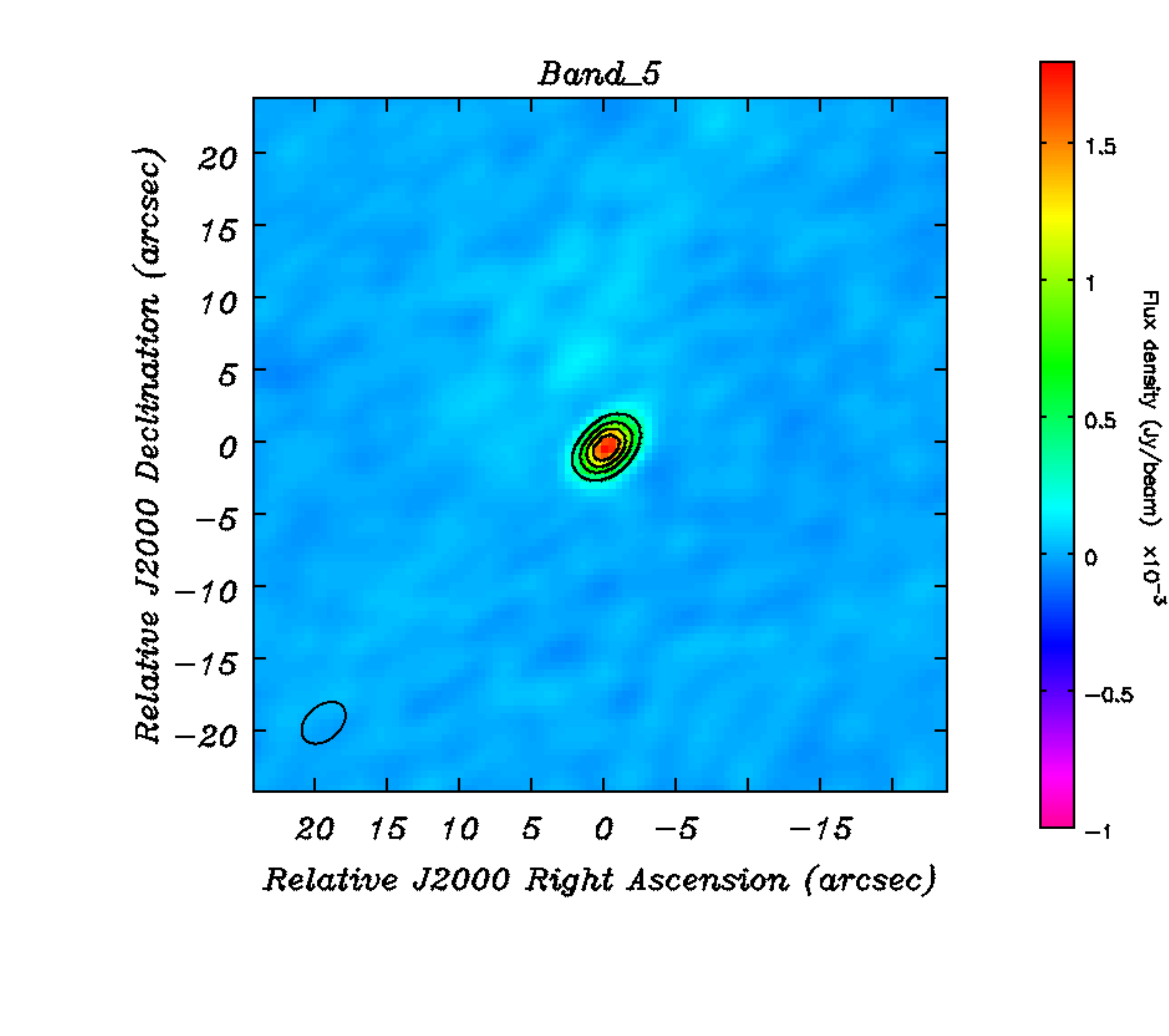}{0.5\textwidth}{(a)}
          \fig{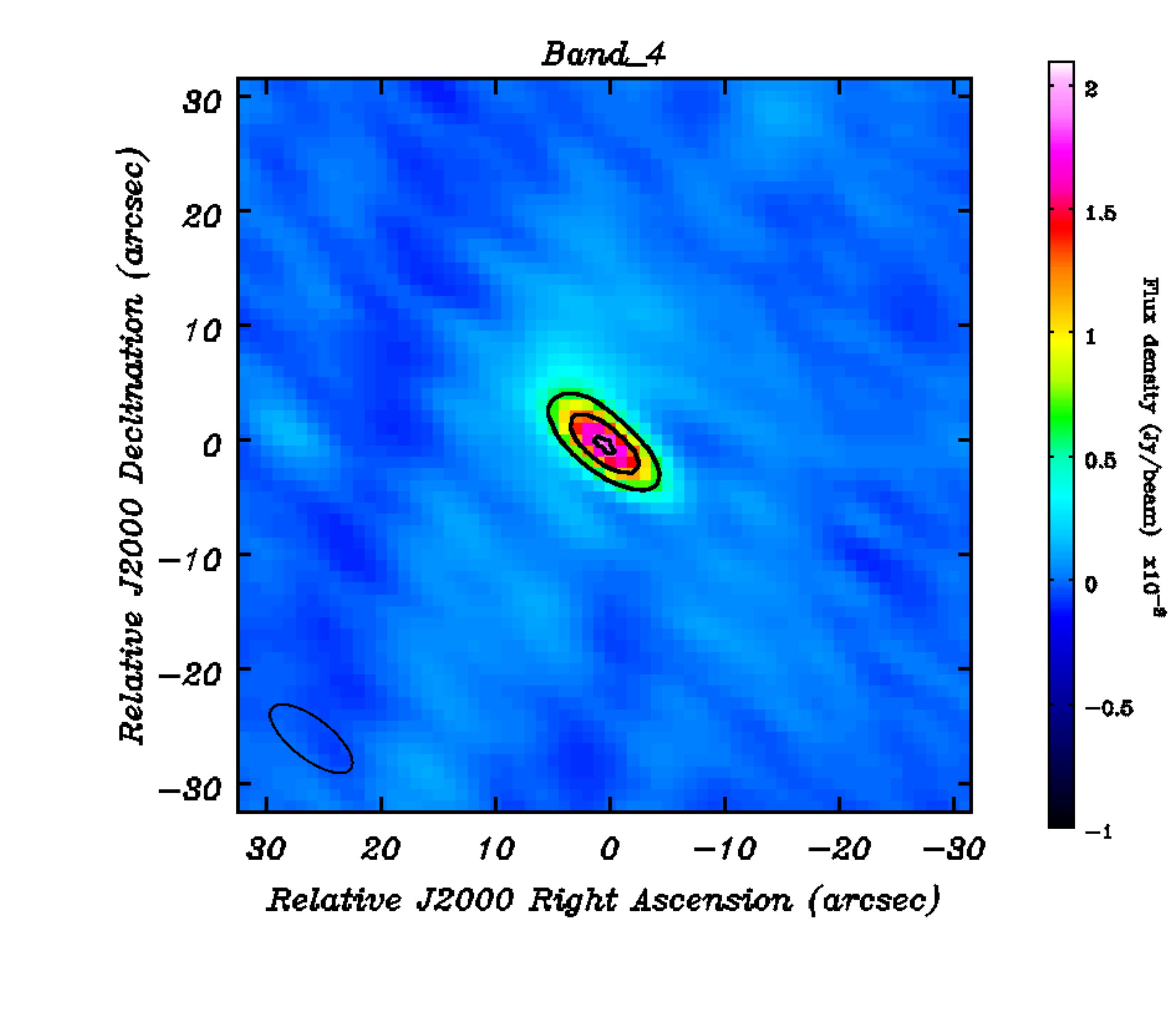}{0.5\textwidth}{(b)}
                    }
\gridline{\fig{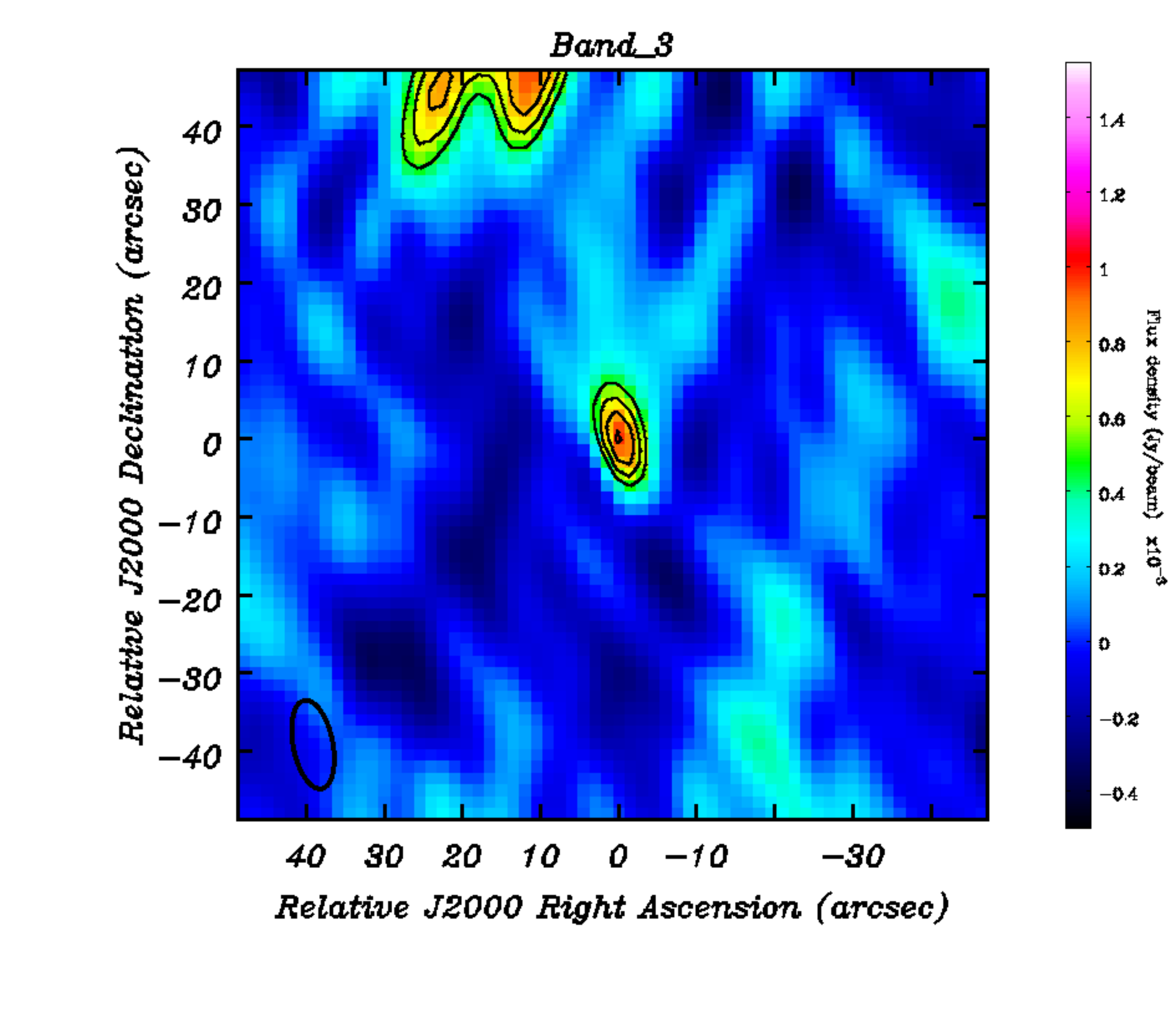}{0.5\textwidth}{(c)}
                    }
\caption{ {  GMRT images of GRB171205A. Panel  a) Band 5 detection on  2019 May 13. Panel b)  Band 4 detection on  2019
September 10.
Panel  c) Band 3 detection on  2019 May 13. The contours in black lines show detection significance and are at  20$\sigma$, 40$\sigma$ and  60$\sigma$ for bands 4 and 5; and at 6$\sigma$, 8$\sigma$, and 10$\sigma$ for band 3, where  $\sigma$ is the map rms of corresponding images. For the images displayed, they are following: Band 5, 17 $\mu$Jy, Band 4, 30 $\mu$Jy and Band 5, 80 $\mu$Jy.}
\label{fig:pyramid}}
\end{figure*}

We use a package Common Astronomy Software Applications( CASA) for data analysis.
The data were analyzed in three major steps, i.e flagging, calibration and imaging.  The CASA task `flagdata' was used to remove dead antennas and bad data. In addition, 
the tasks `tfcrop'(\url{http://www.aoc.nrao.edu/~rurvashi/TFCrop/TFCropV1/node2.html}) and `rflag'(\url{https://casa.nrao.edu/Release4.2.2/docs/userman/UserMansu167.html})  were used to flag the radio frequency interference (RFI). The calibration \citep{wbook} was performed to remove the instrumental and atmospheric effects from the measurement.
The final part of processing was imaging. The continuum imaging of the target source was done using CASA task `tclean'.
Finally, a few  rounds of `phase only' mode and two rounds of `amplitude-phase' self-calibrations were run. We fit a Gaussian to
determine the GRB flux density at the GRB position.
The flux densities are shown in Table \ref{tab:flux}. Sample radio images of GRB171205A at bands 5,
4 and 3 are shown in Figure \ref{fig:pyramid}.
The errors in flux densities in Table \ref{tab:flux} show only the statistical errors. We also add 15\%  of flux densities in quadrature to account the uncertainties due to calibration and other systematics for GMRT bands during our model fit. We closely follow the procedure shown in \citet{pk}.
The table \ref{tab:radio} list the details of observations and flux densities at various epochs.

We also extract {\it Swift}-XRT 0.3--10\,keV flux light curve from the the Swift online 
repository\footnote{\url{https://www.swift.ac.uk/xrt_products/00794972}}. The light curve post day 1 indicate 
 a photon index  $\Gamma=1.94^{+0.23}_{-0.22}$ and column density 
 $N_H=(1.2^{+0.8}_{-0.7} \times 10^{21}$ cm$^{-2}$\footnote{\url{https://www.swift.ac.uk/xrt_live_cat/00794972/}}. 
 This is in addition to Galactic column density of $5.89\times10^{20}$ cm$^{-2}$.
We converted X-ray flux into into 1\,keV spectral flux density using this  photon index.  Swift data covered observations until 2020 May 27.

In addition, we analysed two archival data from {\it Chandra} ACIS-S on 2018 Feb 14 and 2018 June 29 (PI: Margutti).  We used
Chandra Interactive Analysis of Observations software \citep[CIAO;][]{fruscione+06} task
\texttt{specextractor} to 
extract  the spectra, response and ancillary matrices.  We used
CIAO version 4.6 along with CALDB version 4.5.9.  The  HEAsoft\footnote{\url{http://heasarc.gsfc.nasa.gov/docs/software/lheasoft/}}
package Xspec version 12.1 \citep{arnaud+96} was used to carry out the  analysis of the {\it Chandra} spectra.  The GRB was detected in the first obserations at 71 
days with 0.3--10\,keV unabsorbed flux of $(11.25\pm3.38)\times10^{-15}$ erg cm$^{-2}$ s$^{-1}$. The second {\it Chandra} observation on second 
epoch, i.e 206 days, resulted in a 3-$\sigma$ upper limit $< 3.98\times10^{-15}$ erg cm$^{-2}$ s$^{-1}$.

\begin{deluxetable*}{lcccrc}

\tablenum{1}
\tablecaption{Flux Densities of GRB171205A\label{tab:flux}}
\tablewidth{0pt}
\tablehead{
\colhead{Date of Observation} & \colhead{Band} & \colhead{Frequency} & \colhead{Days since explosion} &
\colhead{Flux Density\tablenotemark{a}} & \colhead{Map RMS}  \\
\colhead{} & \colhead{}  & \colhead{MHz} & \colhead{} &
\colhead{(mJy)} & \colhead{$\mu$Jy/beam} 
}
\startdata
2017 Dec 10.10 & 5  & 1255 & 4.79 & $<0.07$ & 24 \\
2017 Dec 11.02 & 4  & 648 & 5.71 & $<$0.06 & 20 \\
2017 Dec 19.91 & 5 & 1265  & 14.60 &$0.64\pm0.05$& 17  \\
2017 Dec 26.90 & 5 & 1265  & 21.59 & 1.00$\pm$0.07& 17 \\
2017 Dec 28.91 & 4 & 607  & 23.60 & $<$0.39& 130 \\
2018 Jan 16.94 & 5 & 1265  & 42.63 &1.75$\pm$0.05& 17  \\
2018 Feb 12.85 & 5 & 1370  & 68.54 &3.04$\pm$0.09& 41  \\
2018 Feb 17.85 & 4 & 607  & 73.54 &1.37$\pm$0.15& 115  \\
2018 Mar 20.68 & 5 & 1352  & 105.37 & 5.79$\pm$0.08& 34\\
%
2018 Jun 08.44 & 3 & 402 &185.13 &$2.94\pm0.41$ & 106\\
2018 Jun 10.68 & 5 & 1255  & 187.37 &3.07$\pm$0.11& 26\\
2018 Jun 11.42 & 4 & 745 &188.11 &$2.46\pm0.49$ & 141\\
 2018 Jul 13.47 & 5 & 1250 &220.16 &3.55$\pm$0.12 &17 \\
 2018 Jul 15.35 & 4 & 610 & 222.04&3.13$\pm$0.18 &66 \\
2018 Jul 23.35 & 3 & 402 & 230.04& $2.30\pm0.22$ &64\\
 2018 Jul 26.57 & 5 & 1265 & 233.26& $2.60\pm0.08$ &26\\
2018 Jul 28.35 & 4 & 750  & 235 &2.92$\pm$0.31 & 62\\
 2018 Aug 24.32 & 5 & 1265 &262.01 & $3.32\pm0.06$ & 33\\
 2018 Aug 25.22 & 4 & 607 &262.91 & $1.06\pm0.16$ & 81\\
2018 Sep 21.35 & 5 & 1265 & 290.04 & $3.03\pm0.10$ & 59\\
2018 Sep 23.16 & 3 & 402 &291.85 & $2.01\pm0.15$ & 63\\
2018 Oct 20.28 & 5 & 1250  & 319 & 3.18 $\pm$ 0.13 & 52\\
2018 Oct 26.09 & 3 & 400  & 324.78 &1.77 $ \pm $ 0.27 & 131  \\
 2018 Oct 26.33 & 4 & 607 & 325.02 & $2.61\pm0.26$ & 143\\
 2018 Dec 21.88 & 3 & 402 & 381.57 & $1.61\pm0.31$ & 49\\
 2018 Dec 22.02 & 5 &   1255&381.71  & $2.83\pm0.06$ & 21\\
2018 Dec 22.14 & 4 & 610&  381.83 & $1.52\pm0.17$ &52 \\
2019 Feb 26.98 &  4 & 610  & 448.67  & 1.52$\pm$ 0.21 & 79\\
 2019 Feb 26.70 & 3 & 402  & 448.39  & 1.05$\pm$ 0.16 & 40\\
2019 Feb 26.87 & 5 & 1250  &448.56   &1.87$\pm$0.06  & 17\\
2019 May 13.49 & 5 & 1250  & 524.18 & 1.74$\pm$0.04  &  17\\
2019 May 13.62 & 3 & 402  & 524.31  &1.43 $\pm$0.26 & 79 \\
2019 May 13.77 & 4 & 607  & 524.46  & 2.09 $\pm$0.18 & 81\\
2019 Sep 10.16 &  5 & 1250  & 643.85  &1.74 $\pm$0.03 & 19\\
 2019 Sep 10.31 & 3 & 402  & 644.00  &1.67 $\pm$ 0.21 & 50\\
2019 Sep 10.43 & 4 & 750  & 644.12  &1.93 $\pm$0.10 & 30\\
 2019 Dec 09.91 & 4 & 647  & 734.60  &1.38 $\pm$0.12 & 37\\
 2019 Dec 10.02 &  5 & 1265  & 734.71  &1.71 $\pm$0.04 & 20\\
  2019 Dec 10.19 & 3 & 402  & 734.88  &1.30 $\pm$ 0.17 & 59\\
 2020 June 26.37 & 4  & 648  & 934.06 &  $1.29\pm0.11$ &   20\\
 2020 June 26.49 &  5 & 1255 & 934.18  & $1.23\pm0.04$ & 21\\
  2020 June 29.44  &  3 & 402 &  937.13 & $1.12\pm0.16$ &  55\\
\enddata
\label{tab:radio}
\tablenotetext{a}{The uncertainties reflect the statistical errors.}
\end{deluxetable*}

\section{Modelling and results}

\subsection{GRB afterglow Model} 
\label{sec:model}

We use external synchrotron model for the GRB afterglow emission, which arises due to the interaction between the GRB outflow and the surrounding CBM \citep{granot}.
As the outflow moves into the CBM,   a  `forward shock' or a `blast wave' shock moving into the CBM, and a `reverse shock'  moving into the ejected outflow are created. These shocks have the ability to accelerate charged particles to relativistic speeds via Fermi acceleration \citep{longair}. 
Radio afterglow emission is expected to be synchrotron emission arising due to these relativistic charged particles in the shocks in
the presence of magnetic fields.

The evolution of the blast wave is `self-similar' \citep{blan}, and the  dynamics  depends only on the density of the CBM  and the blast wave energy. The CBM
is  usually modelled to be one of the two forms, a constant density medium  and a wind like density medium  \citep{li}. 
The number density profile of the ambient medium is usually modelled  as a powerlaw  $n \propto r^{-k}$. 
For constant density case, the parameter 
$k=0$ and $n=n_0$. For wind like case, the mass flows radially outwards at uniform speed and rate from the of GRB progenitor  giving 
  $k=2$; hence for a  mass-loss rate from the progenitor $\dot{M}_W$  and the progenitor wind velocity $V_W$, the density can be
  defined as  \citep{GAO,ch}:
 \begin{equation}
n=\frac{\dot{M}_W}{4\pi r^{2} m_p V_W}=3\times 10^{35} A_* r^{-2}
\end{equation} 
Where $m_p$ is the mass of the proton,  $A_*$ is in units of  $3\times 10^{35}$ cm$^{-1}$ (or $5\times10^{11}$ g\,cm$^{-1}$ for mass-density),
corresponding to $\dot{M}_{W,-5}/V_{W,3}$.
Here $\dot{M}_{W,-5}$ is mass loss rate in $10^{-5}M_{\odot}/\rm{yr}$ and $V_{W,3}=10^3\,\rm{km/s}$.

The GRB wideband afterglow spectrum has several breaks characterized by various characteristic frequencies, namely $\nu_a$  (the  transition from optically thick to thin region, i.e. synchrotron self-absorption (SSA) peak), $\nu_c$ (synchrotron cooling frequency)  and $\nu_m$ (frequency corresponding to minimum injected Lorentz factor). In the  fast cooling regime the frequency ordering is $\nu_m > \nu_c$, while in the  slow cooling regime the ordering is opposite. 
Generally afterglow modelling is done in the slow cooling regime, where the most relevant ordering in radio frequencies in first few days are $\nu_a \le \nu_m \le \nu_c$; and then $\nu_m \le \nu_a \le \nu_c$ at later times
\citep{gv14}. However, there has been evidence for fast cooling in some GRBs with high density environments \citep{Chandra}.
The afterglow spectra evolves as $F_{\nu} \propto \nu^2$ ($\nu<\nu_a$) and $F_{\nu} \propto \nu^{1/3}$ 
($\nu_a < \nu< \min(\nu_m,\nu_c)$) for both fast as well as the slow cooling. In the regime  $\min(\nu_m,\nu_c) < \nu < \max(\nu_m,\nu_c) $, the evolution changes to $F_{\nu} \propto \nu^{-1/2}$
and $F_{\nu} \propto \nu^{-(p-1)/2}$ for fast and slow cooling regimes, respectively, and then evolves as $F_{\nu} \propto \nu^{-p/2}$ for 
$ \nu > \max(\nu_m,\nu_c) $, where $p$ is the usual power law index showing particle number distribution with energy in non-thermal emission case.

In addition to standard afterglow models, shock breakout model too has been favoured for low-luminosity GRBs, where the
 breakout of a shock travelling through the stellar envelope may be responsible for gamma-ray emission.  Due to decreasing density of the 
 stellar matter outwards, the shock breakout velocity increases and may become relativistic. \citet{ns12}
have defined a  ‘relativistic breakout closure relation’ between the breakout energy $E_{\rm bo}$, temperature  $T_{\rm bo}$ and duration
$t_{\rm bo}^{\rm obs}$, i.e. 
$( t_{\rm bo}^{\rm obs}/20 \,\rm s) \sim  (E_{\rm bo}/10^{46}\, \rm erg)^{1/2}  (T_{bo}/50 \, \rm keV)^{-2.68}$. This relation has been found to be followed by several low luminosity GRBs.
For GRB 171205A, the relation gives $\sim 80$\,s, which is roughly 1/3rd of the observed duration. However, the large uncertainties in the 
$E_p$ and $E_{\rm iso}$  \citep{delia+18} do not rule out this model.

\subsection{Inputs from high frequency data}
\label{sec:results}

\begin{figure}
    \centering
         \includegraphics[width=0.48\textwidth]{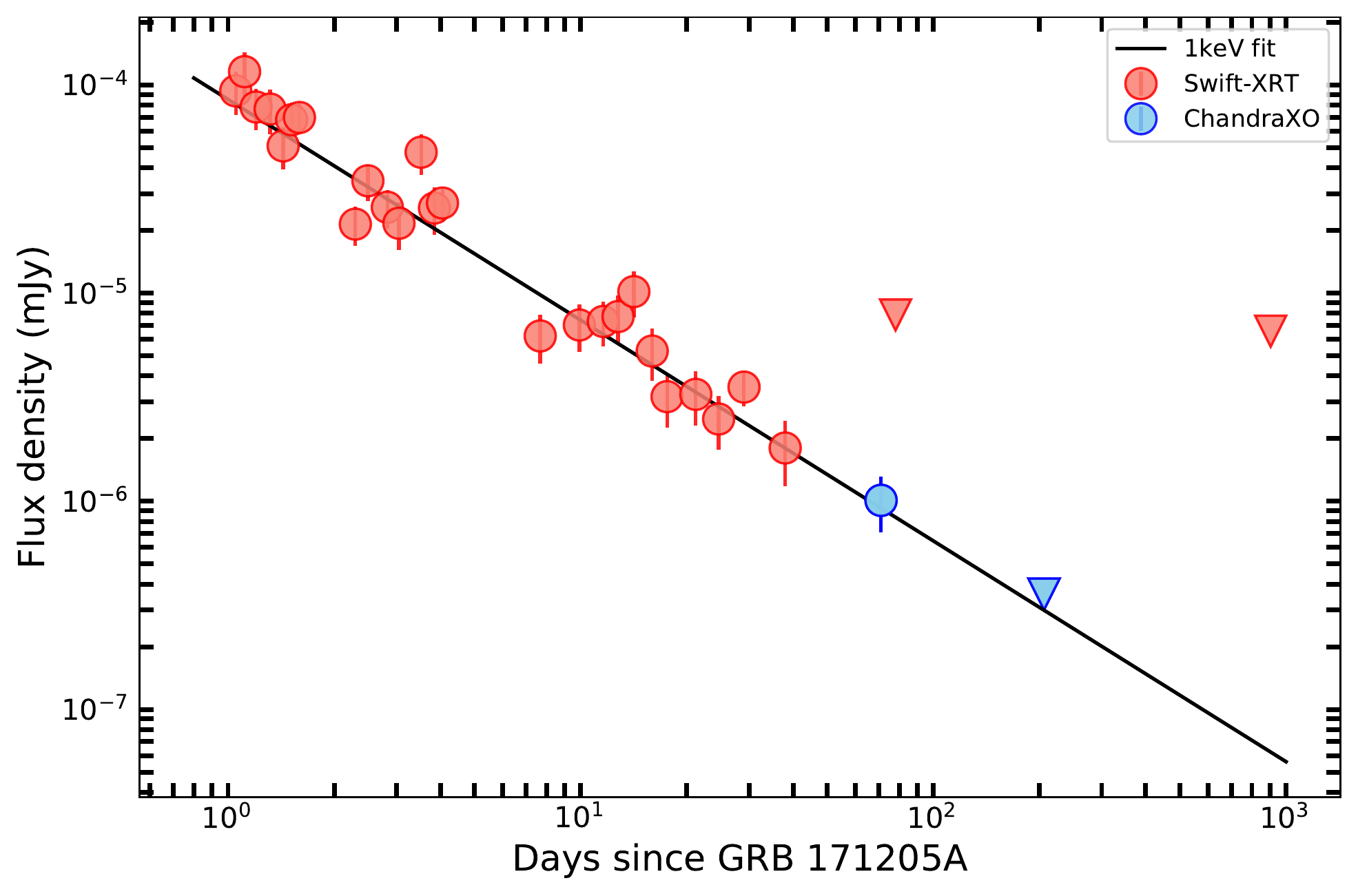}
           \includegraphics[width=0.48\textwidth]{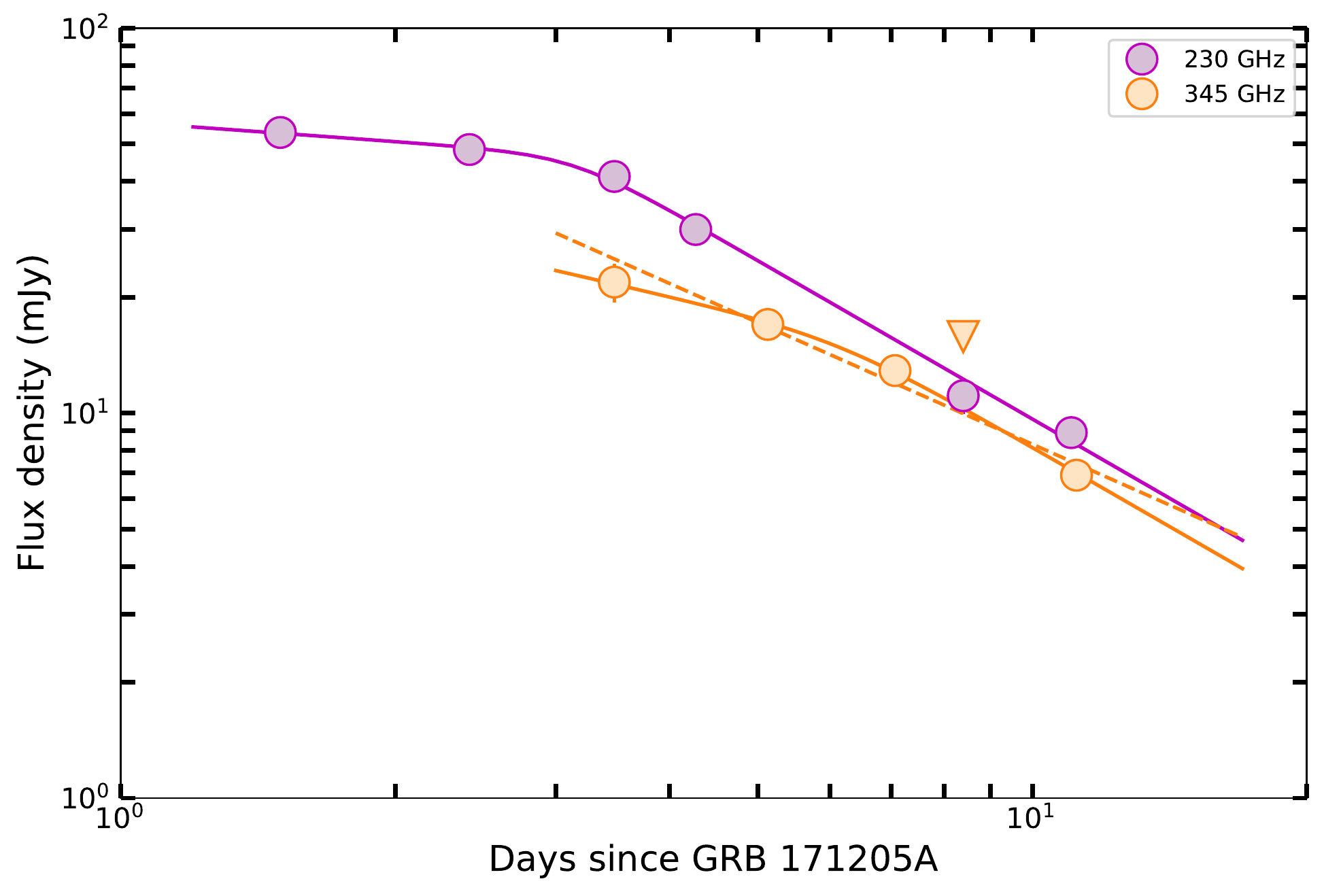}
    \caption{\small{{\it Top panel:}  The 1 keV X-ray light curve of GRB 171205A from day 1 onwards. {\it Swift}-XRT data points are  in red and {\it Chandra} data points are in blue colours. The triangle symbol indicates 3-$\sigma$ upper limit. The light curve is best fit with a 
    powerlaw with an index of $-1.06\pm0.06$. {\it Bottom panel:} Early time 230 and 345 GHz light curves GRB 171205A. The values are taken from \citet{urata}. The light curves are jointly fit with a broken powerlaw with common post peak index (solid lines).  In addition, we also fit the 345\,
    GHz light curve with a simple powerlaw model to determine the significance of the break.
      }}
    \label{fig:xraymm}
\end{figure}

The {\it Swift}-XRT light curve covers epoch until
day $\sim 902$. 
In addition, {\it Chandra} observations are on day 71 and day 205
(Fig. \ref{fig:xraymm}). We fit a power law to the X-ray data post day 1. This is to avoid possible energy injection due to central engine activities at $\le 1$\,d. The light curve is fit with  power law index $\alpha=1.06\pm0.06$. 
\citet{delia+18} also find 
a power law fit with index $1.08$ to the light curve post $\sim$ 1 day,  consistent with
our value. This is typical value for  the  standard X-ray light curve decay before jet break \citep{homo1}, suggesting that
no jet-break was seen until the last {\it Swift}-XRT epoch. The jet break time is thus constrained to
$t_{\rm jet}\ge 71\,\rm d$ \citep{rh}, i.e. the last detected {\it Chandra} epoch.  

The photon index of the {\it Swift}-XRT data is a photon index  $\Gamma=1.94^{+0.23}_{-0.22}$ \footnote{\url{https://www.swift.ac.uk/xrt_live_cat/00794972/}}.  This suggests $\beta=0.94^{+0.23}_{-0.22}$.
The values of $\alpha$ and $\beta$ are consistent with X-ray frequency ($\nu_X$) being above the cooling frequency ($\nu_X>\nu_c$),
 for both wind as well as ISM density profiles. 
The X-ray  temporal index is consistent with $\alpha=(3p-2)/4$ and spectral index $\beta=p/2$. These values gives $p$ to be $2.08\pm0.08$
and $1.88\pm0.46$, respectively, which, within the errorbars,  are consistent with each other.  Within the errorbars, this is also consistent with $\nu_X<\nu_c$ for a ISM like medium where $\alpha=3(p-1)/4$ and $\beta=(p-1)/2$.

We  also plot early time 230 and 345 GHz light curves GRB 171205A, taken from \citet{urata}.   We estimate the realistic errorbars in
the flux density values by adding 5\% of the flux density in quadrature to the map rms to account for the systematic errors.
The mm highest flux density is significantly higher than the peak flux densities at uGMRT bands, indicating most likely presence of wind like medium.
The mm light curves are jointly fit with a broken powerlaw with common 
   post peak index.  The data are best fit with  post break index $-1.37\pm0.07$ (bottom panel of Fig. \ref{fig:xraymm}).
   For above values of $p$, this is consistent with an evolution of $(3p-1)/4$, for $\nu_m<\nu_{\rm mm}<\nu_c$ for wind density profile.
   The pre-break 
   indices $-0.17\pm0.14$ and $-0.54 \pm 0.31$ for 230 and 345 GHz bands, respectively.
    The breaks in
    230 and 345 GHz are at $3.14\pm0.28$ day and $6.12\pm1.17$ day, respectively. 
    At the epochs of the breaks, the flux density of the 230 and 345 GHz light curves are $43.83\pm3.29$ and $14.85\pm 2.86$ mJy, respectively.  We also fit the 345 GHz data with a single powerlaw model to determine the significance of the break. The single powerlaw model fits with an index
    of $-1.05\pm0.14$, however results in a much larger reduced-$\chi^2$ value of 2.47 as compared to 0.81 in the broken power law case. 
    For the broken powerlaw model, the  characteristic frequency 
    evolves with an index of $+0.71\pm0.12$., hence the breaks
    can not be due to passage of $\nu_m$, which decreases with time.  
    The only possibility is the breaks are being due to $\nu_c$ in the wind medium, where $\nu_c \propto t^{1/2}$.  
    Thus the data indicate. $\nu_c\approx 230$ GHz on day 3.14 and $\nu_c\approx 345$ GHz on day 6.12.  
           However, there is  one concern. The pre-break evolution is rather flat. This could be reconciled if  mm bands are close to passage of $\nu_m$. 
           Another possibility is that if there is a reverse shock component which is contributing to the mm band at the early epoch.   
    The  thick shell reverse shock model during the reverse shock crossing phase will evolve as $-(p-2)/2$, for 
    $\nu_{\rm mm}>\nu_c$ in  the slow cooling phase and  $\nu_{\rm mm}>\nu_m$ in fast cooling phase.
    The  characteristic frequency 
    evolves with an index of $+0.71\pm0.12$, this is in between evolution of the $\nu_c\, (\propto t^{1/2}) $ in the  forward shock      
     and the evolution of 
    $\nu_c\, (\propto t^{1}) $ in the reverse shock. This also suggests the possibility of reverse shock contributing to the afterglow model.  If so,  then 
    the breaks at day 3.14 and 6.12 in 230 and 345 GHz are artificial breaks and may not reflect the cooling break.

\begin{figure*}
    \centering
    \includegraphics[width=0.3\textwidth]{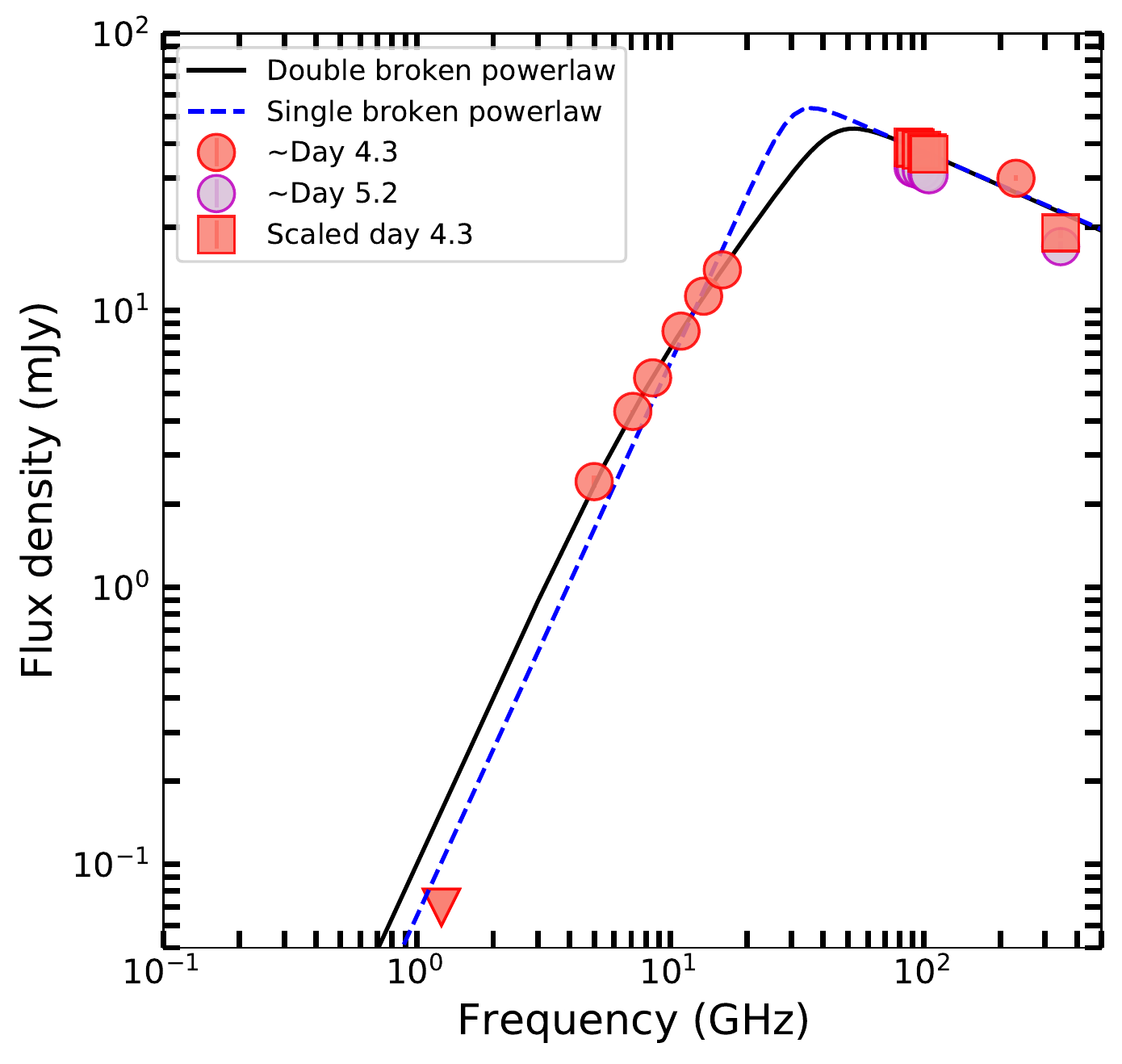}
    \includegraphics[width=0.3\textwidth]{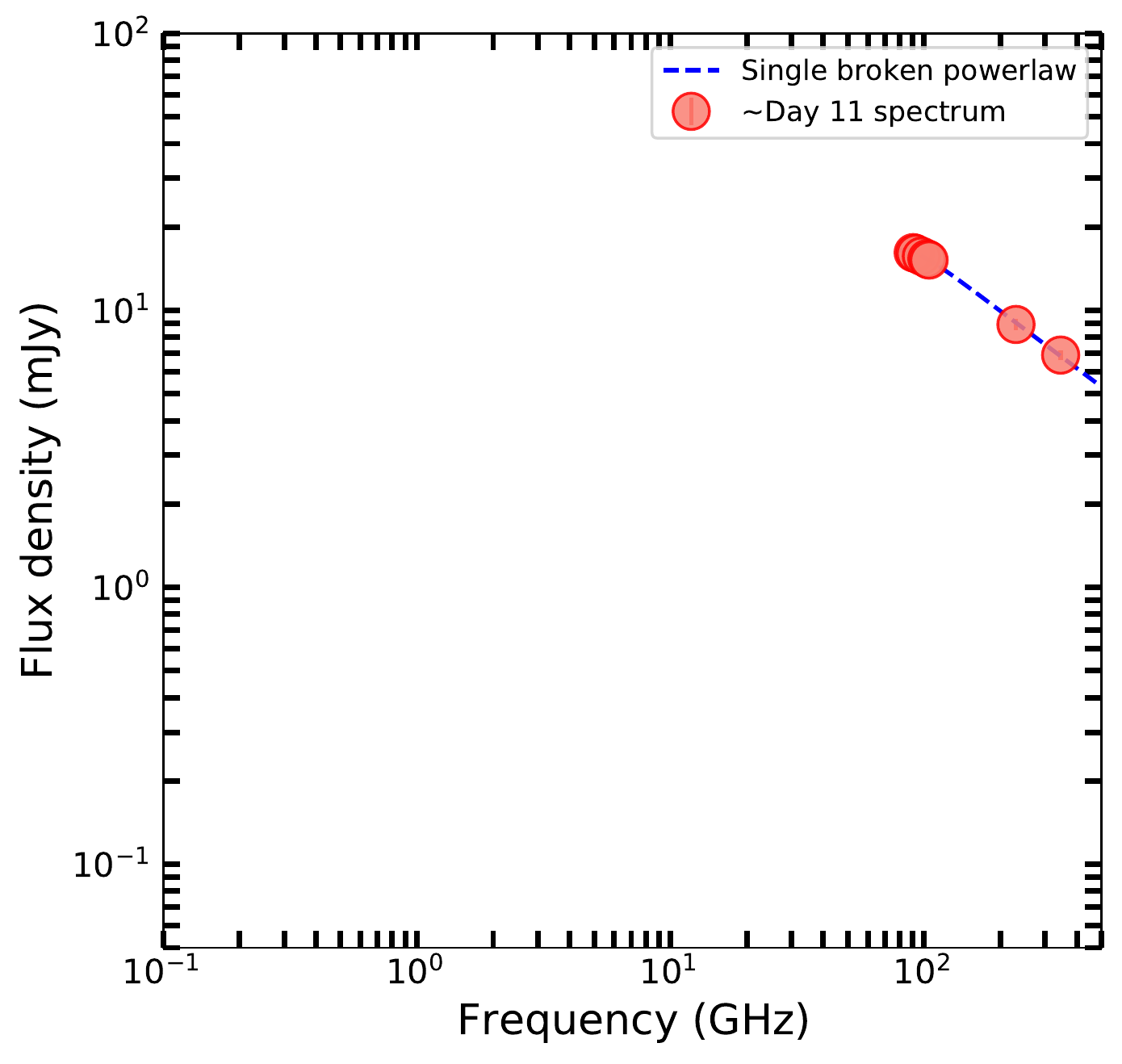}
          \includegraphics[width=0.3\textwidth]{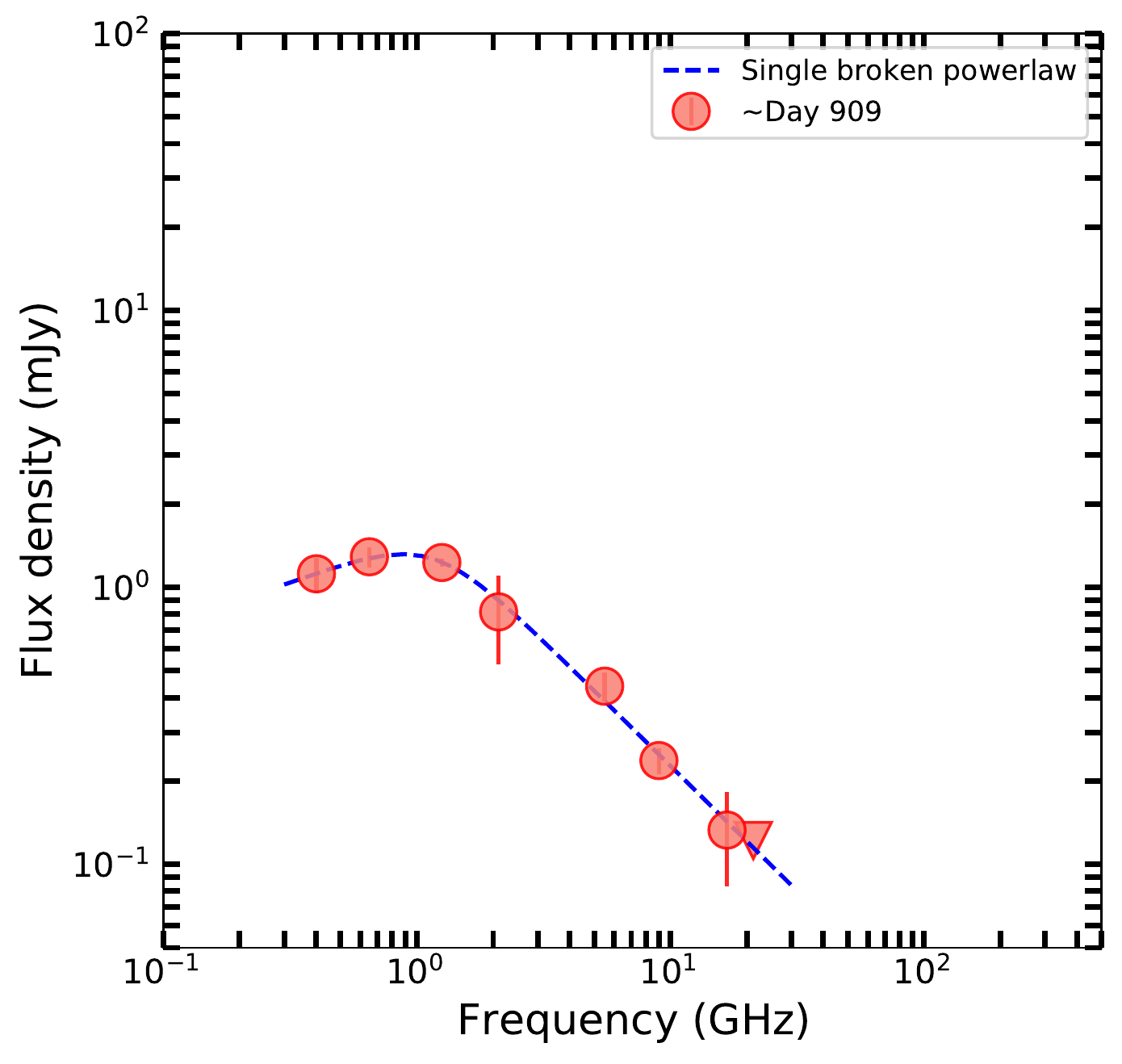}
              \includegraphics[width=0.3\textwidth]{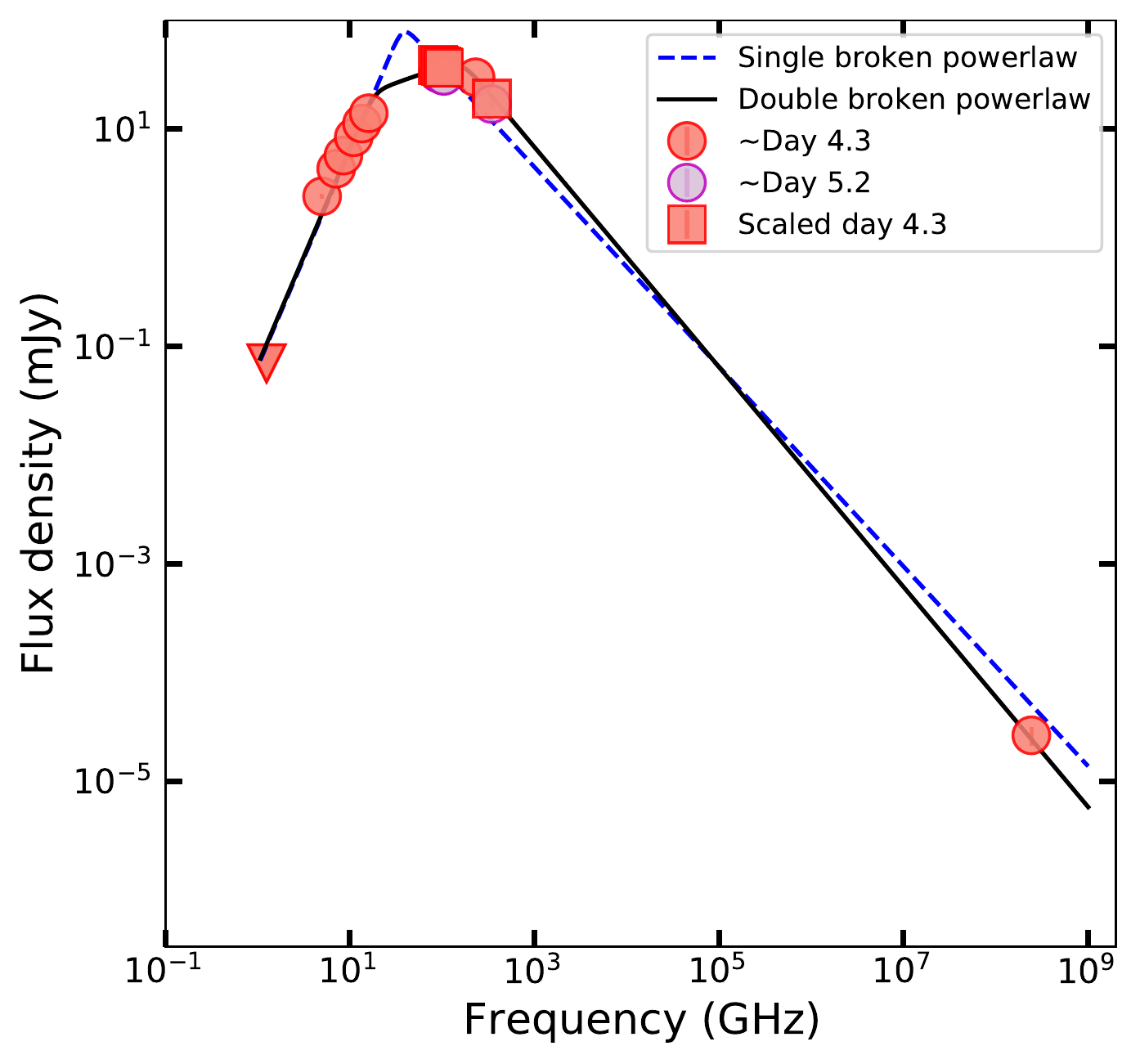}
    \includegraphics[width=0.3\textwidth]{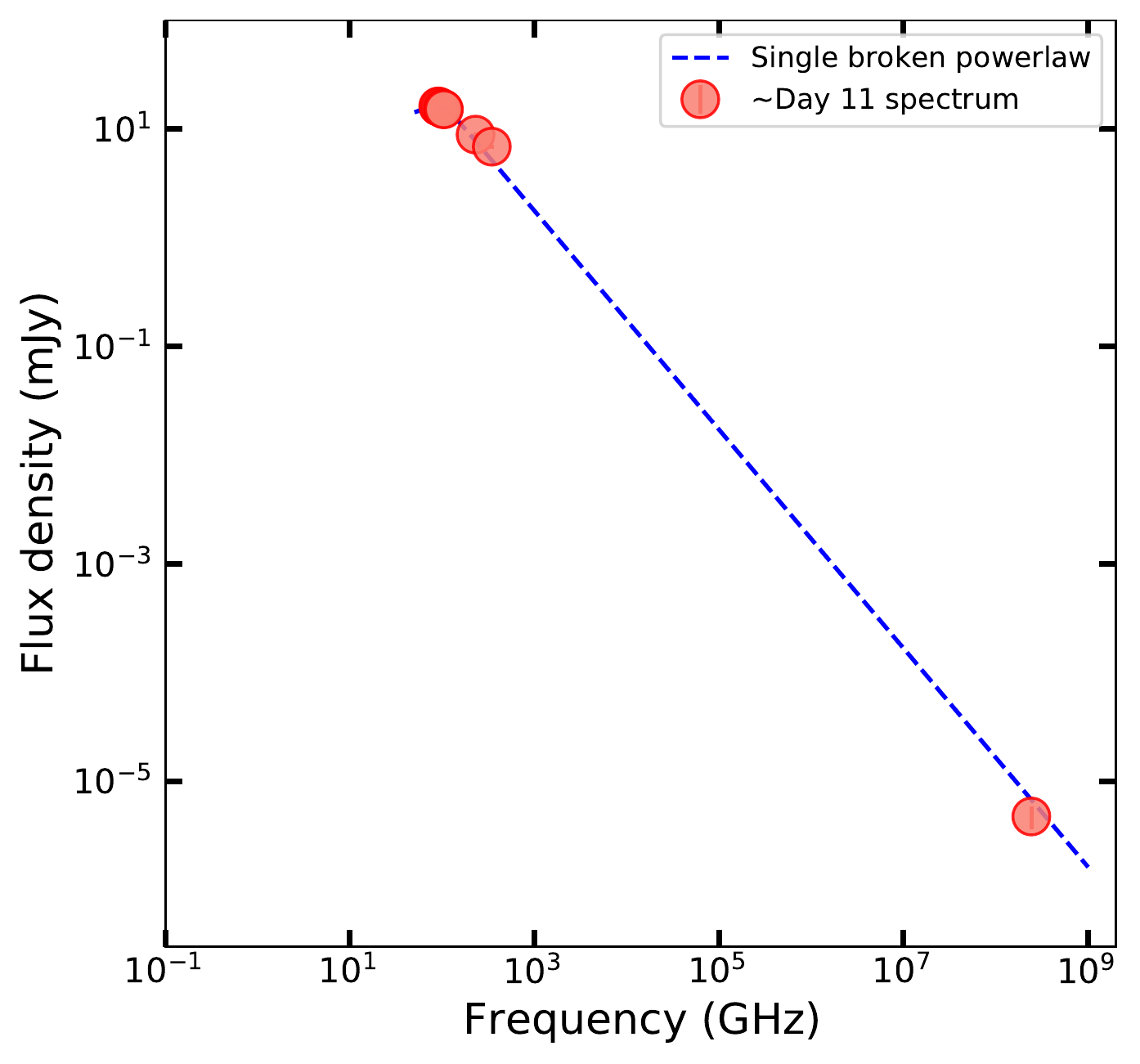}
       \includegraphics[width=0.3\textwidth]{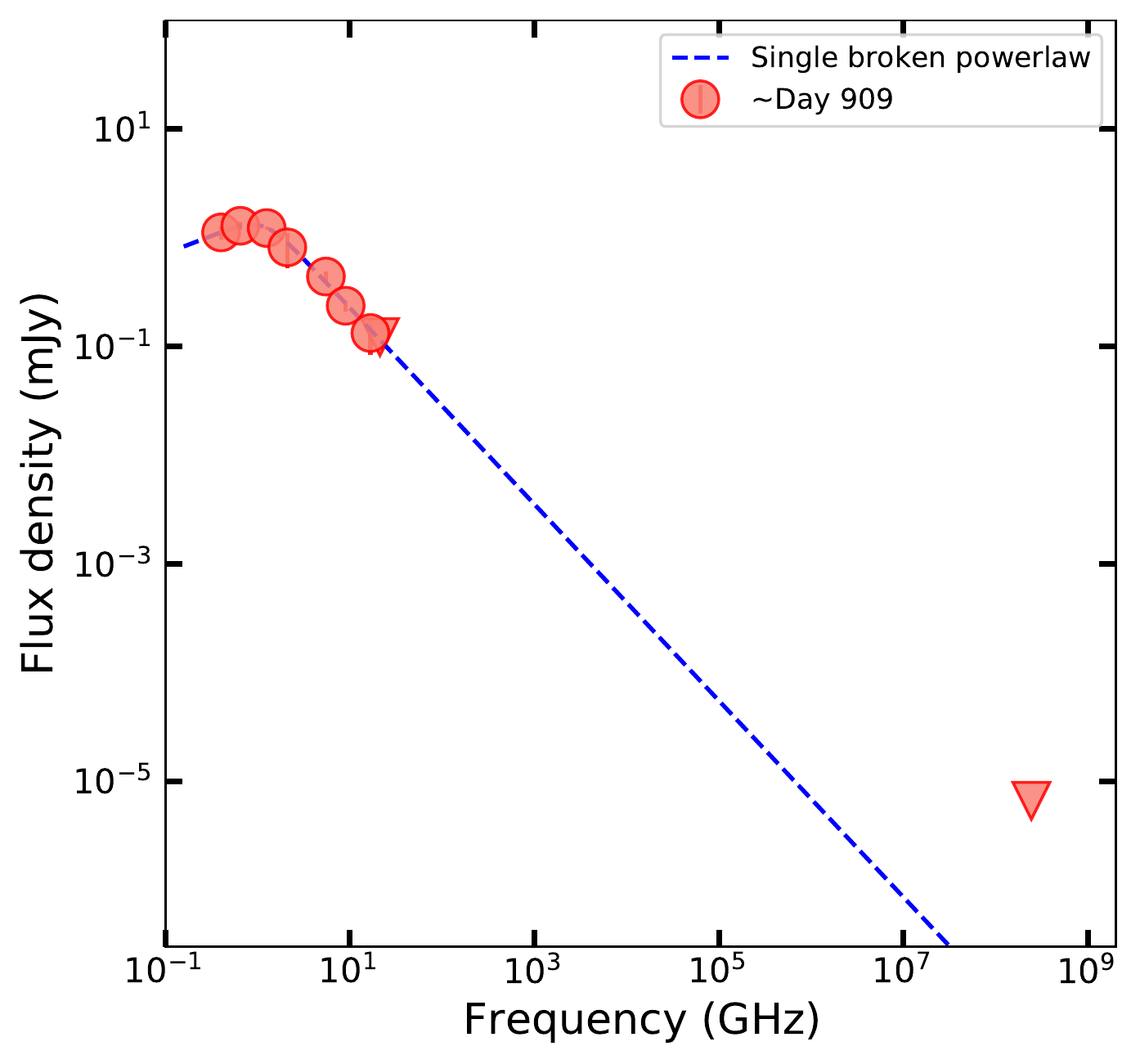}
    \caption{\small{{\it Top panel:} Near-simultaneous spectra on day $\sim 4$, day $\sim 11$ and day $\sim 909$. Here we use only the cm and mm
  data. {\it Bottom panel:} The same as top panel but including the X-ray data as well. The spectra are fit with smoothed single broken powerlaw fits. For day 4.2, we also fit the data with 
   a powerlaw with two breaks. The double broken powerlaw fit is indicated with a black continuous line and the single one with dashed blue line.
   }}
    \label{fig:spec}
\end{figure*}

We also combine the early epoch published cm and mm data from \citet{urata}, late epoch Australian Telescope Compact Array (ATCA) data from
\citet{leung+20} with the uGMRT and the X-ray data, to  obtain near simultaneous spectra on around day 4, day 11 and day 909
(Fig. \ref{fig:spec}). For the 
early epoch spectrum, the VLA data are on day 4.3 and the ALMA data are on day 5.2. We use the nearest epoch temporal evolution to derive the 
ALMA values on day 4.3. In the figure, we plot the values on  day 5.2 as well as their derived values on day 4.3 along with the VLA data. 
We do not use  optical data as supernova signatures appeared by day 3 and
hence the optical data are      likely to be heavily contaminated by
the underlying supernova. 
 We fit a smoothed single broken powerlaw (SBPL) and smoothed double broken powerlaw (DBPL) fits to day 4.3 spectrum.
  We used formalism of \citet{granot} for the treatment of smoothening of  powerlaws at the break frequencies. 
 We first fit only the cm and mm data.
For day 4.3, the best fit SBPL model peaks at $30.41\pm4.08$ GHz and afterwards evolves as $-0.41\pm0.25$.  Here we have fixed the pre-break spectral index to 2. However, even when we use this as a free parameter, the best-fit index is consistent with 2 within errorbars. The 
DBPL model fits the data well and give the breaks at $7.25\pm2.11$ GHz and $44.41\pm3.62$ GHz with post break indices $1.26\pm0.12$, and 
$-0.41\pm0.02$, respectively.
 The peak flux density is $43.80\pm1.12$ mJy.
  
  We also fit SBPL for the spectrum on day 11 and 909.
  The data indicates a fit $-0.68\pm0.02$ with peak $<90$ GHz, with the peak flux density $>16.2$ mJy. The pre-break index is $0.43\pm0.46$.
  The spectrum on day 909 are fit with 
pre-break and post-break indices of $0.15\pm0.13$ and  $-0.90\pm0.04$, respectively, and a peak at  $1.55\pm0.48$ GHz.
The peak flux at day 909 is $1.15\pm0.16
$ mJy.

Now we carry out the above fits including the X-ray data as well. The data to mm to X-ray data are  fit by indices close to $-1$.
These values are  $-1.01\pm0.18$
and $-1.01\pm0.02$ for the first two spectra, respectively.  For the day 909, the X-ray upper limit do not
constrain the model. The mm to X-ray indices  consistent with the X-ray spectral indices between 0.3--10\,keV. 
 This indicates that the cooling frequency ($\nu_c$) is probably close to mm values if $ \nu_X > \nu_c$. Due to lack of optical data, we cannot constrain the cooling frequencies more precisely.

The peak flux density and the frequency of the peak in the three cases are $37.79\pm3.95$ mJy at $160.93\pm35.31$ GHz, 
$>16.2$ mJy at   $<90$ GHz and $1.15\pm0.16$ mJy at $1.55\pm0.48$ GHz., respectively.
Our analysis indicate that the peak of the spectra on day 4 and day 909 are due to $\nu_m$ or $\nu_a$.
Between day 4 and 909, the peak flux density evolves as $-0.63\pm0.02$.
This clearly rules out ISM model and supports the wind model. 

 \citet{zhang+07}  has estimated kinetic energy in the synchrotron afterglow,  $E_K$,  from the X-ray data at the time of shallow to normal decay, 
 which for GRB 171205A is 1.05 day.
For $\nu_X>\nu_c$,  $E_K$ is independent of density and is  only weakly depends on $B$ and $p$, and therefore an ideal regime to measure $E_K$. One can then derive $E_K$ from the X-ray band using  Eq 9 of \citet{zhang+07}, which in this case is 
$E_K\approx 1.4\times 10^{50} (\epsilon_B/0.01)^{2-p/2+p} (\epsilon_e/0.1)^{4(1-p)/2+p}$\, erg.

\subsection{Initial inferences from uGMRT data}

\begin{figure}
    \centering
   \includegraphics[width=0.48\textwidth]{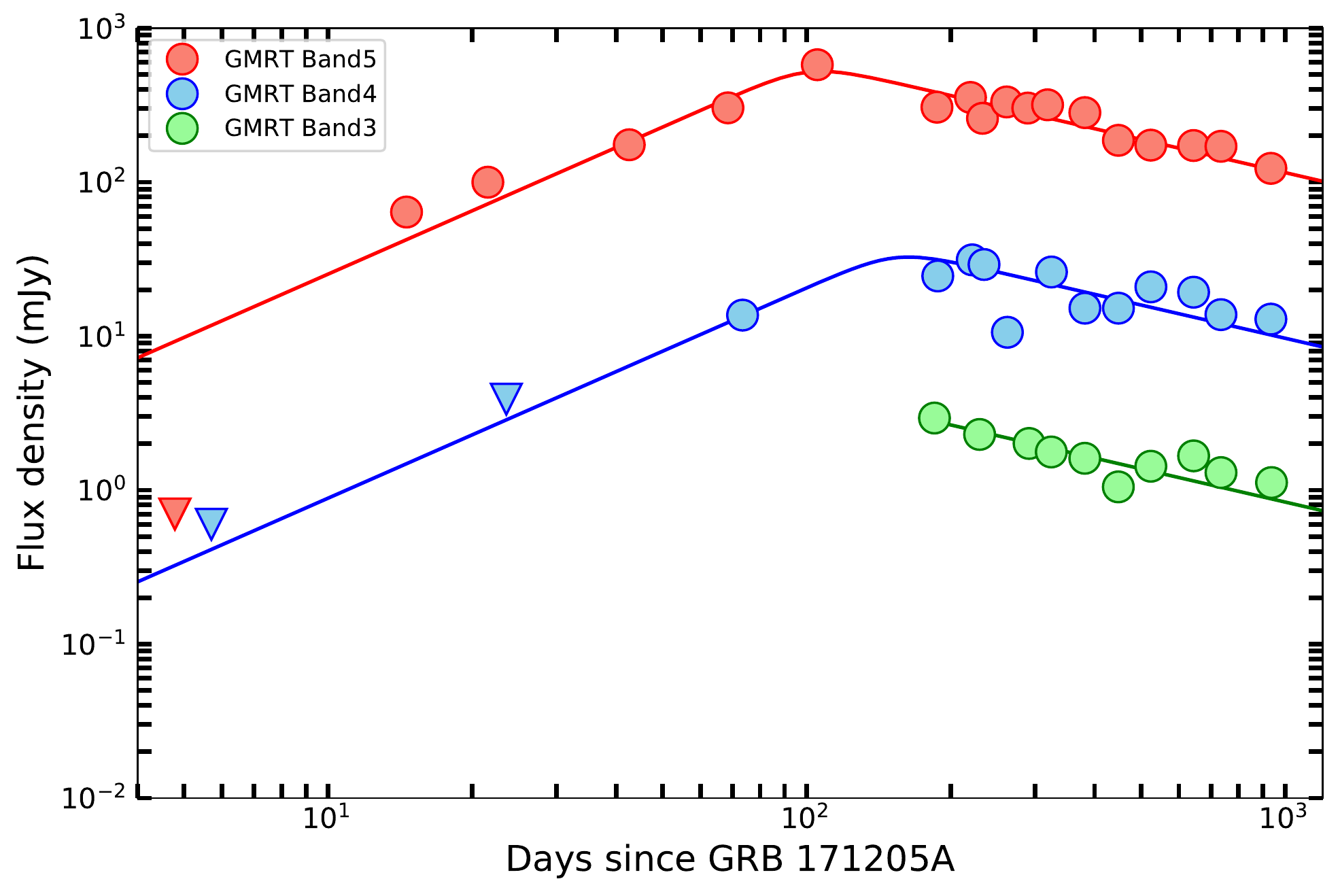}
    \caption{{\small{The uGMRT bands 5, 4 and 3 radio light curves (the band 4 and 5 values are scaled by factors of 10 and 100). The data are best fit with pre- and post peak spectral indices of $1.37\pm0.20$ and $-0.72\pm0.06$.}}}
    \label{fig:radio}
\end{figure}

 We plot uGMRT radio light curves and fit them jointly with a smoothed broken powerlaw (SBPL) model. We allow the normalization to vary but 
fix the indices before and after the peak.   We show the light curves in Fig. \ref{fig:radio}.
In the figure, the band 4 and 5 values are scaled by factors of 10 and 100, respectively, for clarity.
The indicies before and after the peak are
$1.37\pm0.20$ and $-0.72\pm0.06$.

The evolution before peak at uGMRT frequencies ($\nu_{\rm radio}$) is in rough agreement with wind slow cooling case for 
$\nu_{\rm radio}<\nu_a$  for
$\nu_a< \rm min(\nu_m,\nu_c)$; the wind fast cooling phase for $\nu_a<\nu_c<
\nu_m$ if the observing frequency is in the transition zone between $\nu_{\rm radio}<\nu_a$ to $\nu_a<\nu_{\rm radio}<\nu_c$, as well as
wind slow cooling phase for $\nu_m<\nu_a<
\nu_c$ for transition between $\nu_{\rm radio}<\nu_m$ to $\nu_m<\nu_{\rm radio}<\nu_a$.  The post peak index is rather shallow and is consistent only with the 
wind fast cooling in the regime $\nu_a<\nu_{\rm radio}<\nu_c$. It is rather shallow for the post jet-break or non-relativistic evolution. However, as we discuss in the next section, the shallow decline of radio light curve in seen in other GRBs as well, and other emission components may contribute to it.

From our fits, the epochs of the peak flux densities are 
$5.19\pm0.05$ and $3.23\pm0.15$ mJy, respectively, in bands 5 and 4
on day $101.08\pm9.58$ and day $153.71\pm40.31$, respectively. 
In band 3, the data are optically thin, which constraints the peak to be $<185.13$ day and 
flux
$>2.81$ mJy. 
 The peak flux density evolves as $-1.13 \pm  0.76$. While this value has a large error, it is consistent with the evolution in the stratified wind within 2-$\sigma$ and
most likely rule out all the models involving ISM, where
$F_{\rm max} \propto t^0$. This also rules out $1<p<2$ case for which $F_{\nu, \rm max}$  is
expected to remain constant \citep{GAO}.

\subsection{Model fits}

In this section, we carry out detailed model fits to the uGMRT data.
The simple closure relations seem to suggest that the GRB is in a slow cooling regime  with wind density medium.
We fit the data with all three scenarios, i.e. 
wind density slow cooling regime $\nu_a<\nu_m<\nu_c$, $\nu_m<\nu_a<\nu_c$ and wind density fast cooling
regime $\nu_a<\nu_c<\nu_m$. We also account for the non-standard
wind profile, i.e. $k \neq 2$.  
For this we keep $k$ as a free parameter and  adopt the expressions shown in \citep{thesis}.

Models are proposed for different stages of blastwave expansion. The precise coefficients associated with specified model parameters can be computed by numerical simulations. For decelerating blastwave  and adiabatic wind like case the parameter dependencies of peak flux density
and characteristic frequencies are given by \citep{GAO}
\begin{align}
F_{\nu, \rm max} & = 17.0\,\rm{mJy} \left(\frac{1+z}{2}\right)^{3/2}\left(\frac{E_{\rm KE}}{10^{49}\, {\rm erg}}\right)^{1/2}   \left(\frac{\epsilon_{B}}{0.01}\right)^{1/2}\nonumber \\
 & \left(\frac{A_*}{0.1 \dot{M}_{W,-5}/V_{W,3} }\right) \left(\frac{D}{100\,\rm Mpc}\right)^{-2} \left(\frac{t}{10\,\rm d}\right)^{-1/2} 
\end{align} 
\begin{align}
\nu_c  & =1.7\times10^{17}\,\rm{Hz}\left(\frac{1+z}{2}\right)^{-3/2}\left(\frac{E_{\rm KE}}{10^{49}\, {\rm erg}}\right)^{1/2}  \nonumber \\
& \left(\frac{A_*}{0.1 \dot{M}_{W,-5}/V_{W,3} }\right) ^{-2}\left(\frac{\epsilon_{B}}{0.01}\right)^{-3/2}\left(\frac{t}{10\,\rm d}\right)^{1/2}
\end{align}
\begin{align}
\nu_m & =2.7\times10^{7}\,\rm{Hz} \,G'(p) \left(\frac{1+z}{2}\right)^{1/2}\left(\frac{E_{\rm KE}}{10^{49}\, {\rm erg}}\right)^{1/2}  \nonumber \\
&  \left(\frac{\epsilon_{e}}{0.1}\right)^{2}
\left(\frac{\epsilon_{B}}{0.01}\right)^{1/2}\left(\frac{t}{10\,\rm d}\right)^{-3/2}
\end{align}
here  the parameters $\epsilon_B$ and $\epsilon_e$ are microscopic parameters indicating fraction of energy into magnetic field and relativistic electrons, respectively and $E_{\rm KE}$ is the afterglow kinetic energy.
 $G'(p)=0.053 ((p-2)/(p-1))^2$.

The expression $\nu_a$ for the three regimes are:
\begin{equation}
\nu_a
\begin{cases} 
=4.3\times10^{9}\,{ \rm Hz}\, g'(p) \left(\frac{1+z}{2}\right)^{-2/5}   \left(\frac{E_{\rm KE}}{10^{49}\, {\rm erg}}\right)^{-2/5} \\  
\left(\frac{A_*}{0.1 \dot{M}_{W,-5}/V_{W,3} }\right) ^{6/5}  
\left(\frac{\epsilon_{B}}{0.01}\right)^{1/5} \left(\frac{\epsilon_{e}}{0.1}\right)^{-1}   \left(\frac{t}{10\,\rm d}\right)^{-3/5}
 , & \\ \text{for} \ \nu_a<\nu_m<\nu_c. \\ 
 =4.9\times10^{8}\,{ \rm Hz}\, g''(p) \left(\frac{1+z}{2}\right)^{\frac{p-2}{2(p+4)}}  \left(\frac{E_{\rm KE}}{10^{49}\, {\rm erg}}\right)^{\frac{p-2}{2(p+4)}}\\  
\left(\frac{A_*}{0.1 \dot{M}_{W,-5}/V_{W,3} }\right) ^{\frac{4}{p+4}}  
\left(\frac{\epsilon_{B}}{0.01}\right)^{\frac{p+2}{2(p+4)}} \left(\frac{\epsilon_{e}}{0.1}\right)^{\frac{2(p-1)}{p+4}}  \left(\frac{t}{10\,\rm d}\right)^{-\frac{3(p+2)}{2(p+4)}}
 , & \\ \text{for} \ \nu_m<\nu_a<\nu_c.\\ 
 =6.0\times10^{4}\,{ \rm Hz} \, g'''(p) \left(\frac{1+z}{2}\right)^{3/5}\left(\frac{E_{\rm KE}}{10^{49}\, {\rm erg}}\right)^{-2/5} \\  
\left(\frac{A_*}{0.1 \dot{M}_{W,-5}/V_{W,3} }\right) ^{11/5}  
\left(\frac{\epsilon_{B}}{0.01}\right)^{6/5}\left(\frac{t}{10\,\rm d}\right)^{-8/5}
 , & \\ \text{for} \ \nu_a<\nu_c<\nu_m. \\ 
\end{cases}
\end{equation}
Here are expressions for $g'(p)$, $g''(p)$ and $g'''(p)$ are derived in \citet{GAO} and for $p=2.1$.
Using these expressions, the temporal and spectral evolution in different transitions regimes can be derived and are
mentioned in  \citet{GAO}.

 We also carry out modeling for the shock breakout cases, for which 
we adopt methodology of \citet{bd+15}. In this model,
due to decreasing outer ejecta density, the  outer parts of the shock envelope are faster and less energetic, and inner parts are slower and more energetic.  As slower material catches up with the decelerating ejecta it re-energizes the forward shock 
and  the blastwave  energy  continuously changes with time.   Thus this model can be treated as a series of successive shells 
which accelerate and catch up to the boundary and hence explain the increasing afterglow energy via continuous injection \citep{bd+15}.
If $\eta$ is the ratio of the prompt to afterglow energy ($\eta \equiv  E_{\rm \gamma.iso}/ E_{\rm k,iso})$, then for shock breakout case, we
 parameterize the model as 
$\eta_{\rm eff} E_{\rm k,iso} = E_{\rm \gamma.iso}(\frac{tA_*}{1+z})^s$, where $s$ is free parameter characterising energy injection  \citep{bd+15}.

Using this expression along with the scalings for $\nu_a$, $\nu_m$, $\nu_c$ and $F_{\rm max}$ for various regimes, provided by \citet{bd+15},
leads to the following closure relations for a wind like medium:

Case 1 ($\nu_a<\nu_m<\nu_c$):\\

\begin{equation}
F_{\nu}(t) \propto
\begin{cases} \nu^2 t^{s+1}, & \text{for} \ \nu<\nu_a \\ 
\nu^{1/3} t^{\frac{s}{3}}, & \text{for} \ \nu_a<\nu<\nu_m \  \\
 \nu^{-(p-1)/2} t^{-\frac{3p-1 -s(p+1)}{4}} , & \text{for} \ \nu_m<\nu<\nu_c \\ 
\nu^{-p/2} t^{-\frac{3p-2 -s(p+2)}{4}}, & \text{for}\ \nu>\nu_c
\end{cases}
\end{equation}

Case 2 ($\nu_m<\nu_a<\nu_c$):\\

\begin{equation}
F_{\nu}(t) \propto
\begin{cases} \nu^2 t^{s+1}, & \text{for} \ \nu<\nu_m \\ 
\nu^{5/2} t^{\frac{7+3s}{4}}, & \text{for} \ \nu_m<\nu<\nu_a \  \\
 \nu^{-(p-1)/2} t^{-\frac{3p-1 -s(p+1)}{4}} , & \text{for} \ \nu_a<\nu<\nu_c \\ 
\nu^{-p/2} t^{-\frac{3p-2 -s(p+2)}{4}}, & \text{for}\ \nu>\nu_c
\end{cases}
\end{equation}

Case 3 ($\nu_a<\nu_c<\nu_m$):\\

\begin{equation}
F_{\nu}(t) \propto
\begin{cases} \nu^2 t^{2+s}, & \text{for} \ \nu<\nu_a \\ 
\nu^{1/3} t^{-\frac{2(s+1)}{3}}, & \text{for} \ \nu_a<\nu<\nu_c \  \\
 \nu^{-1/2} t^{-\frac{s+1}{4}} , & \text{for} \ \nu_c<\nu<\nu_m \\ 
\nu^{-p/2} t^{-\frac{3p-2 +(p+2)s}{4}}, & \text{for}\ \nu>\nu_m
\end{cases}
\end{equation}

We now fit the uGMRT data with both standard isotropic afterglow and shock break-out  afterglow models. 
We use  smoothed broken powerlaw models for various regimes following the recipe of  \citet{granot}.
We use $z=0.036$, $D=163$\,Mpc. We define 
$E_{\rm KE}=E_{\rm \gamma.iso}/\eta$, and keep
$\eta$ ($\eta_{\rm eff}$ for SBO) as the free parameter. 
The parameters $p$, $A^*$, $\epsilon_B$  and  $\epsilon_e$  are also  free parameters.

With the inputs above, we carry out the detailed modelling using Markov chain Monte Carlo (MCMC) fitting  using the Python package emcee \citep{fm+13}. We choose 150 walkers, 2000 steps.  Even though the analytical modelling suggests wind like medium, we still start with fits to a constant density medium. The fit results in a high values of reduced-$\chi^2$ further ruling out the constant density model. 
We fit the
standard wind model with $k=2$ for all the cases. In addition, we also account for non-standard wind density medium keeping $k$ as a
free parameter. 

 Table \ref{tab:table} shows the fit statistics for different parameters using the above mentioned models. 
For $\nu_a<\nu_m<\nu_c$ case, we do not list general $k$
 model as this model performed quite poorly for both afterglow as well as shock breakout. 
 $\nu_a<\nu_m<\nu_c$ generally performs very poorly, with shock breakout model performing slightly better than
 the isotropic afterglow model. In addition, the parameters obtained in this model are rather unphysical.
 While the fast cooling model gives best reduced $\chi_{\nu}^{2}$, this case is unlikely to be true.
The analysis of mm and  X-ray light published data  have already revealed that $\nu_c$ lies between the mm and X-ray  frequencies. 
Since $\nu_c \propto t^{1/2}$ in the wind model, uGMRT radio frequencies cannot be in the fast cooling regime.

The most viable model fits are obtained for $\nu_m<\nu_a<\nu_c$ case.  This is quite viable since in wind density provide,
$\nu_m$ evolves faster than $\nu_a$ and may reach $\nu_m<\nu_a$ regime at late epochs 
\citep{gv14}. Here keeping $k$ as free parameter also results in $k \sim 2$. 
Our model fits are equally good for the 
 standard afterglow model and the shock breakout model, and uGMRT data alone cannot differentiate between the two.

\begin{figure*}
    \centering
       \includegraphics[width=0.50\textwidth]{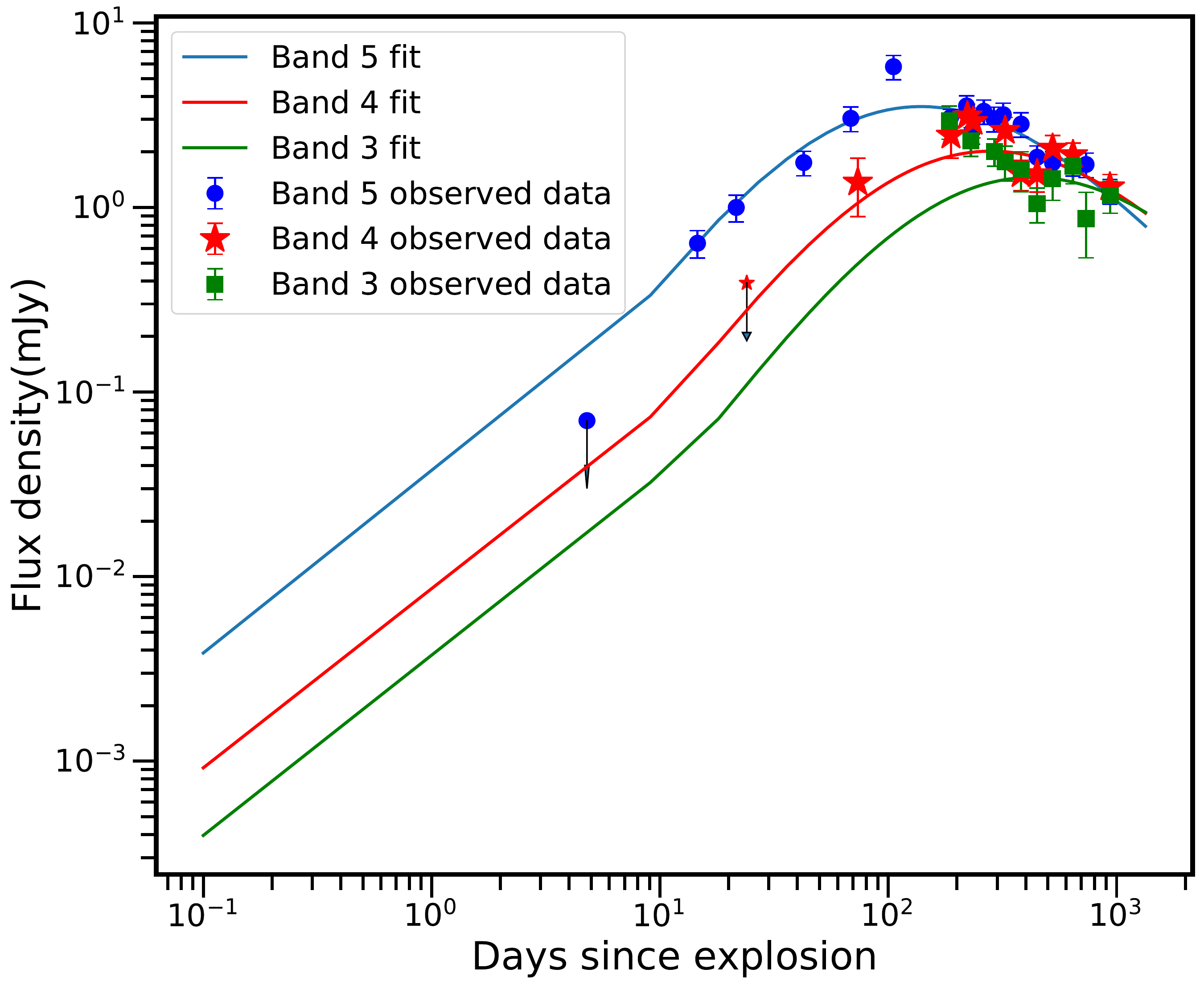}
         \includegraphics[width=0.48\textwidth]{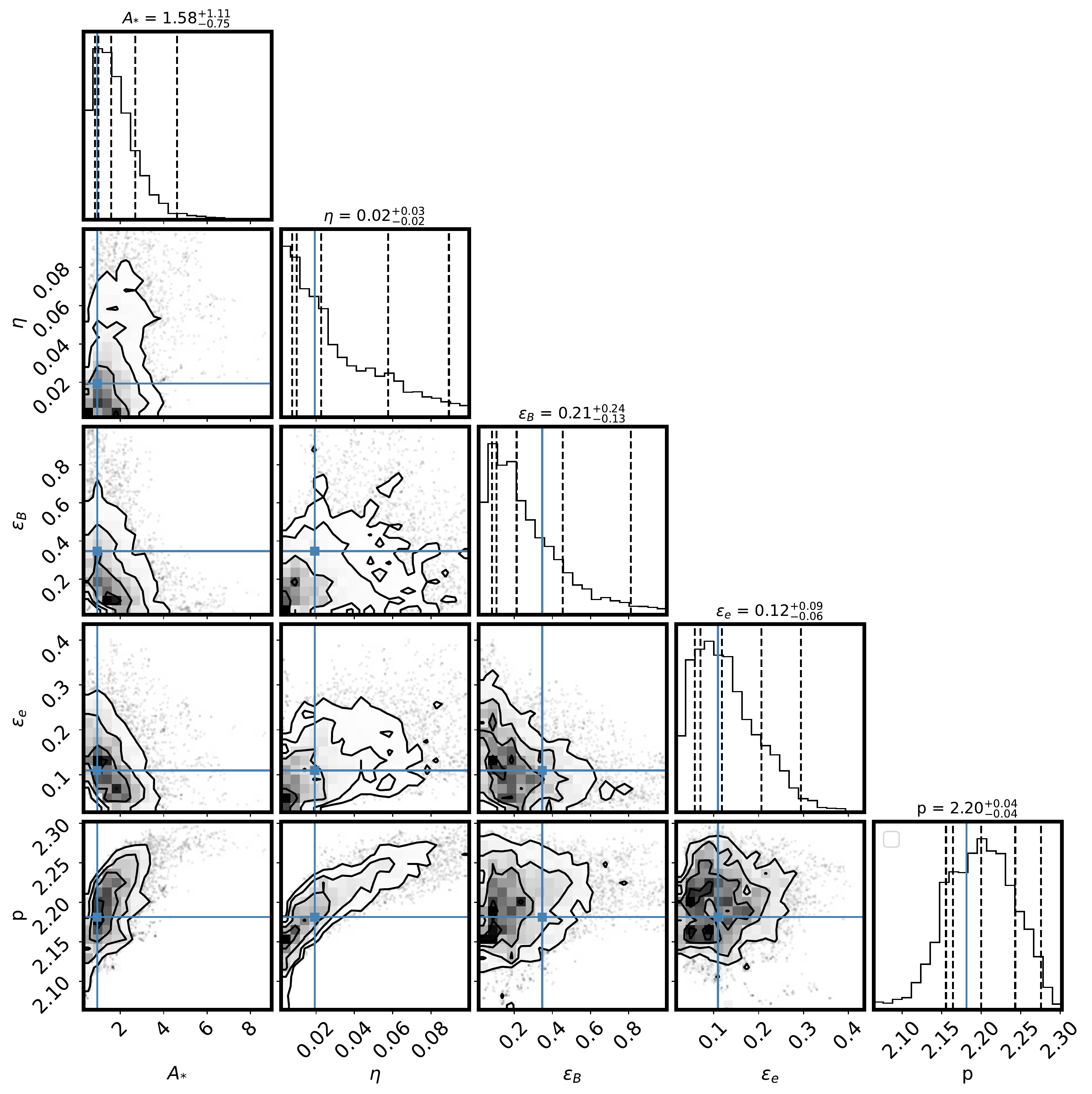}
              \includegraphics[width=0.50\textwidth]{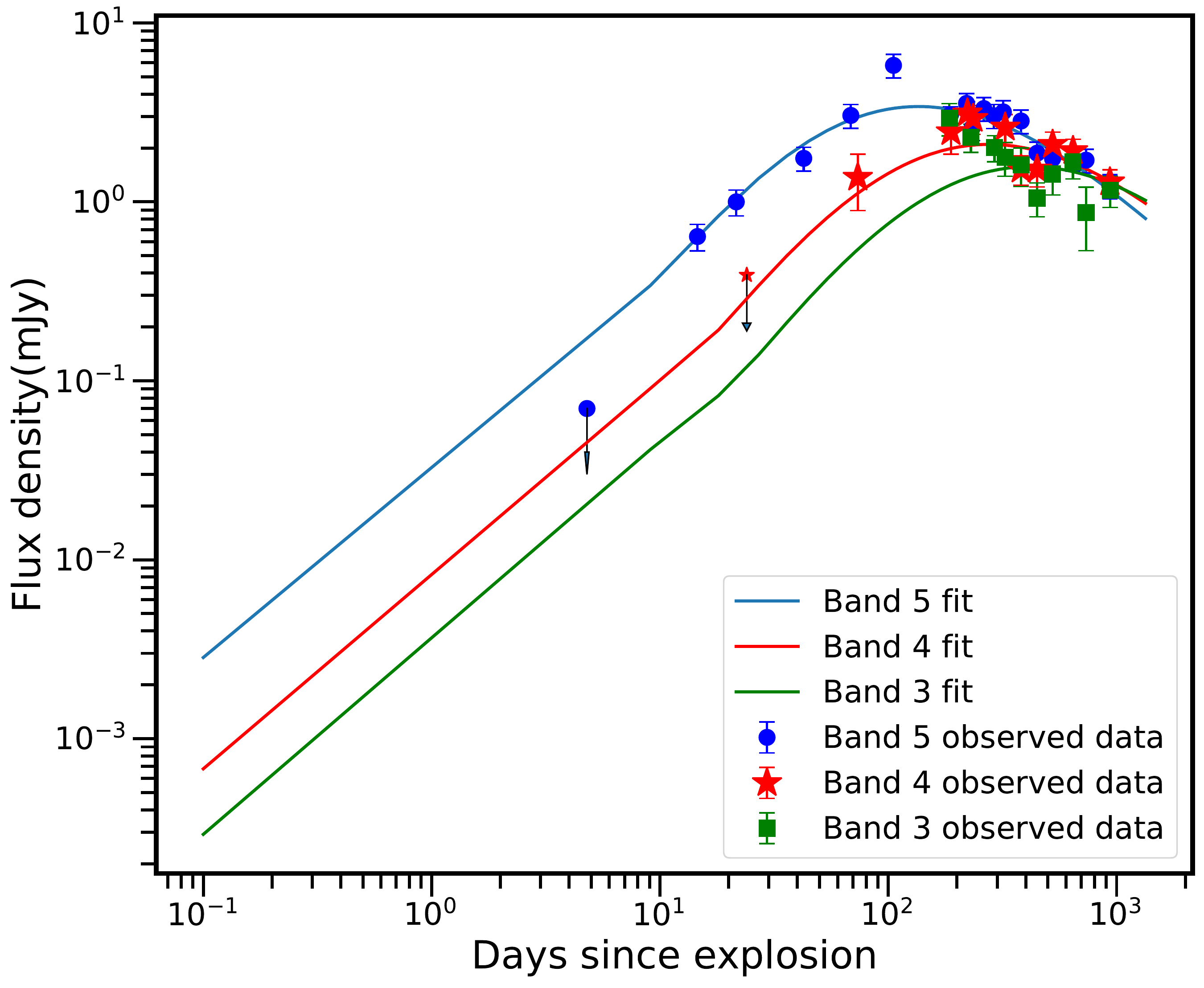}
         \includegraphics[width=0.48\textwidth]{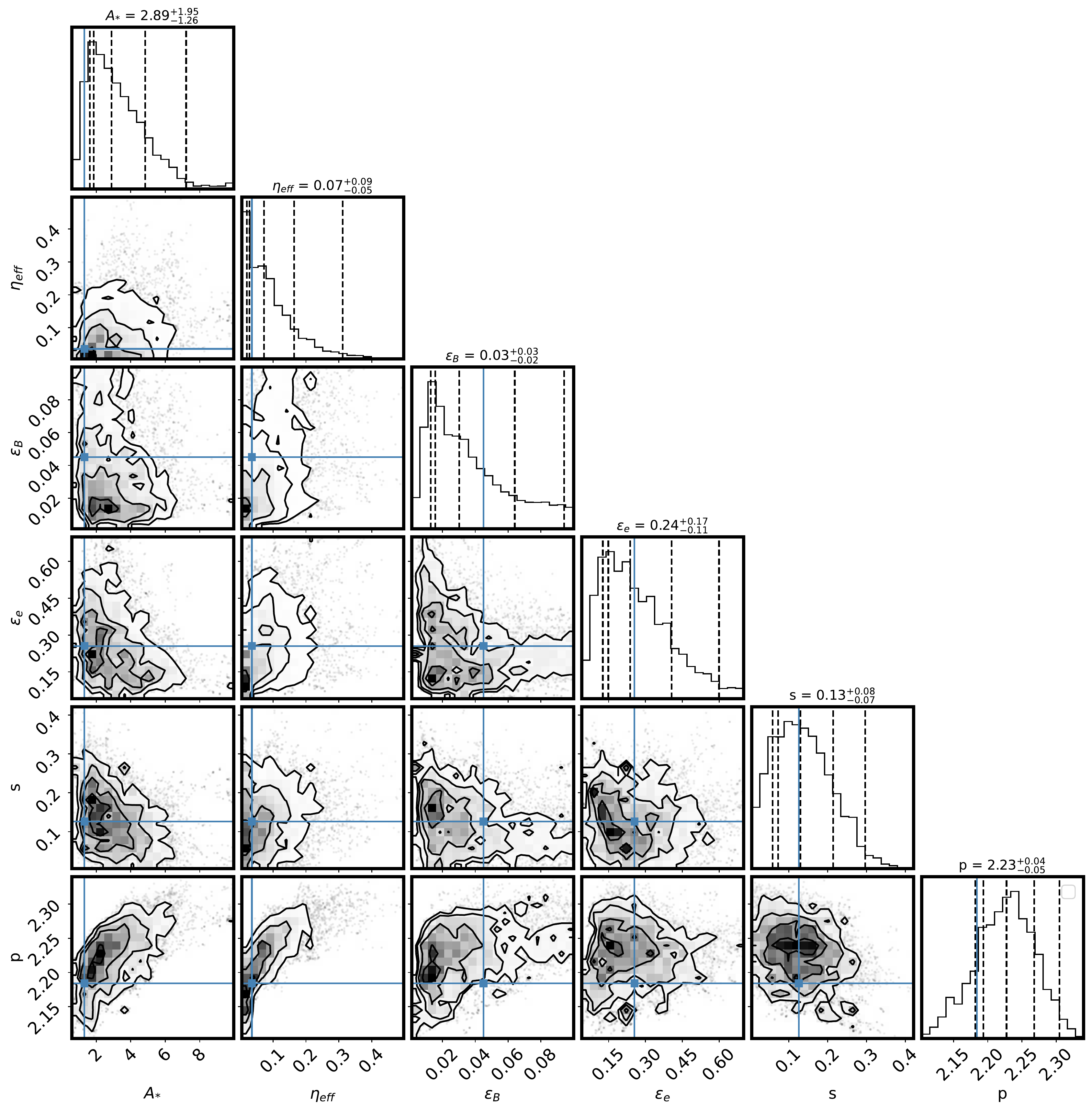}
   \caption{{\small{ Upper left panel: Light curves of GRB171205A in radio regime using the slow cooling model ($\nu_m<\nu_a<\nu_c$) and standard wind ($k=2$).  Upper right panel: Posterior Distributions of parameters for this model.   Lower left panel: Light curves of GRB171205A in radio regime using the slow cooling model ($\nu_m<\nu_a<\nu_c$) with standard wind ($k=2$) and shock breakout scenario.  Lower right panel: Posterior Distributions of parameters for this model. 
   In the left panels, the  blue, red and green points are observed data in band 5 , band 4 and band 3 respectively which are included in the fit. The  points with arrow have only the upper limit of flux. The lines are the best fits. In the right panels, the 2D plots show the joint probability distribution of any two parameters. The contours are at 0.5$\sigma$, 1$\sigma$, 2$\sigma$, 3$\sigma$. The middle dotted lines in the 1D parameter distribution is the median value of posterior followed by 1$\sigma$ and 2$\sigma$  lines on both sides. $\sigma$ is standard deviation of the corresponding distribution.}} }  
      \label{fig:radiofits2}
\end{figure*}

 In Fig. \ref{fig:radiofits2}, we  show the light curve  for a standard wind $k=2$ model for $\nu_m<\nu_a<\nu_c$ case for both isotropic 
 forward shock afterglow
 as well as the shock breakout afterglow model.

\begin{longrotatetable}
\begin{deluxetable*}{|l|ll|llll|llll|}
\tablecaption{Best fit parameters for GRB 171205A uGMRT data  \label{tab:table}}
\tablecolumns{9}
\tablewidth{0pt}
\tablehead{
\colhead{Param.} & \multicolumn{2}{|c|}{$\nu_a<\nu_m<\nu_c$} & \multicolumn{4}{|c|}{$\nu_m<\nu_a<\nu_c$} & \multicolumn{4}{|c|}{$\nu_a<\nu_c<\nu_m$}\\
\cline{2-3}
\cline{4-7}
\cline{8-11}
& \colhead{AG }  & \colhead{SBO} & \multicolumn{2}{|c|}{AG} & \multicolumn{2}{|c|}{SBO}  & \multicolumn{2}{|c|}{AG} & \multicolumn{2}{|c|}{SBO}\\
& \colhead{$k=2$}  & \colhead{$k=2$} & \colhead{ $k=2$} & \colhead{general $k$} &
 \colhead{ $k=2$} & \colhead{general $k$} & \colhead{ $k=2$} & \colhead{general $k$} & \colhead{ $k=2$} & \colhead{general $k$}\\
}
 \startdata 
 $A_*$ & $7.37^{+0.95}_{-0.80}$ &  $2.82^{+0.46}_{-0.39}$ & $1.58^{+0.11}_{-0.75}$ & $1.11^{+1.03}_{-0.54}$ &  
 $2.89^{+1.95}_{-1.26}$ &  $3.54^{+2.46}_{-1.55}$ & $1.69^{+1.15}_{-0.51}$ & $0.17^{+0.19}_{-0.09}$ &
 $2.15^{+1.53}_{-0.78}$ & $0.22^{+0.16}_{-0.11}$ \\
 $\eta(\eta_{\rm eff}$ for SBO) & $0.005^{+0.0004}_{-0.0004}$ &  $0.30^{+0.13}_{-0.12}$ & $0.02^{+0.02}_{-0.01}$ & $0.02^{+0.03}_{-0.02}$ & $0.07^{+0.09}_{-0.05}$ &
 $0.13^{+0.12}_{-0.07}$ & $0.014^{+0.002}_{-0.002}$ &  $0.03^{+0.003}_{-0.003}$ & $0.06^{+0.05}_{-0.03}$ & 
 $0.03^{+0.02}_{-0.01}$\\
$\epsilon_B$ & $ 0.94^{+0.05}_{-0.09}$ &  $0.70^{+0.20}_{-0.20}$ &  $0.21^{+0.24}_{-0.13}$ & $0.11^{+0.17}_{-0.06}$ & $0.03^{+0.03}_{-0.02}$ & 
$0.01^{+0.01}_{-0.01}$ & $0.17^{+0.14}_{-0.10}$  & $0.02^{+0.01}_{-0.01}$ & $0.08^{+0.09}_{-0.05}$ & $0.01^{+0.01}_{-0.01}$ \\
$\epsilon_e$ & $0.99^{+0.01}_{-0.02}$ & $0.78^{+0.16}_{-0.23}$ & $0.12^{+0.09}_{-0.06}$ & $0.13^{+0.11}_{-0.07}$ & $0.24^{+0.17}_{-0.11}$ &
$0.12^{+0.07}_{-0.06}$ &  $\cdots$ & $\cdots$   & $\cdots$   & $\cdots$    \\ 
$p$ & $2.55^{+0.08}_{-0.06}$ & $3.87^{+0.10}_{-0.19}$ & $2.22^{+0.04}_{-0.04}$ & $2.18^{+0.14}_{-0.09}$ &
$2.23^{+0.04}_{-0.05}$ & $2.22^{+0.16}_{-0.13}$ & $\cdots$ & $\cdots$   & $\cdots$   & $\cdots$    \\
$k$ & $\cdots$  & $\cdots$   & $\cdots$   & $1.99^{+0.01}_{-0.02} $& $\cdots$ & $1.99^{+0.01}_{-0.01} $ &  $\cdots$ &  $1.91^{+0.02}_{-0.01} $
& $\cdots$ & $1.92^{+0.02}_{-0.02} $ \\
 $s$ & $\cdots$   & $0.55^{+0.04}_{-0.04}$ & $\cdots$    & $\cdots$    & $0.13^{+0.08}_{-0.07}$ & $0.15^{+0.08}_{-0.08}$ &  $\cdots$& 
 $\cdots$   &  $0.35^{+0.13}_{-0.16}$ & $0.08^{+0.16}_{-0.18}$ \\
 \hline
 & $\chi_{\nu}^{2}= 7.55$  &   $\chi_{\nu}^{2}=2.58$ &          $\chi_{\nu}^{2}=1.74$  & $\chi_{\nu}^{2}=1.80$  &
$\chi_{\nu}^{2}=1.72$ &   $\chi_{\nu}^{2}=1.78$ &          $\chi_{\nu}^{2}=1.60$  & $\chi_{\nu}^{2}=1.52$ &          $\chi_{\nu}^{2}=1.58$  & $\chi_{\nu}^{2}=1.57$  \\
\enddata
\tablecomments{Here AG is the standard isotropic afterglow model and SBO is the shock breakout model. 
In case of fast cooling, the data are only in the regime $\nu_a$ to $\nu_c$ ( 2 to 1/3 transition of spectra), which we don't have $p$ dependencies in 
temporal or spectral slopes.  The only $p$ dependency is in the expression of $\nu_a$ via that ratio of $G(p)$ parameter which we have taken to be 
of order unity for $p\sim 2.1$. The $\epsilon_e$ dependency is also not there as  $\nu_m$  is unconstrained. }
\end{deluxetable*}
\end{longrotatetable}

\section{\textbf{Discussions}}
\label{sec:discussion}

\subsection{Properties of GRB 171205A from radio modelling}

\begin{figure}
\includegraphics[width=0.49\textwidth]{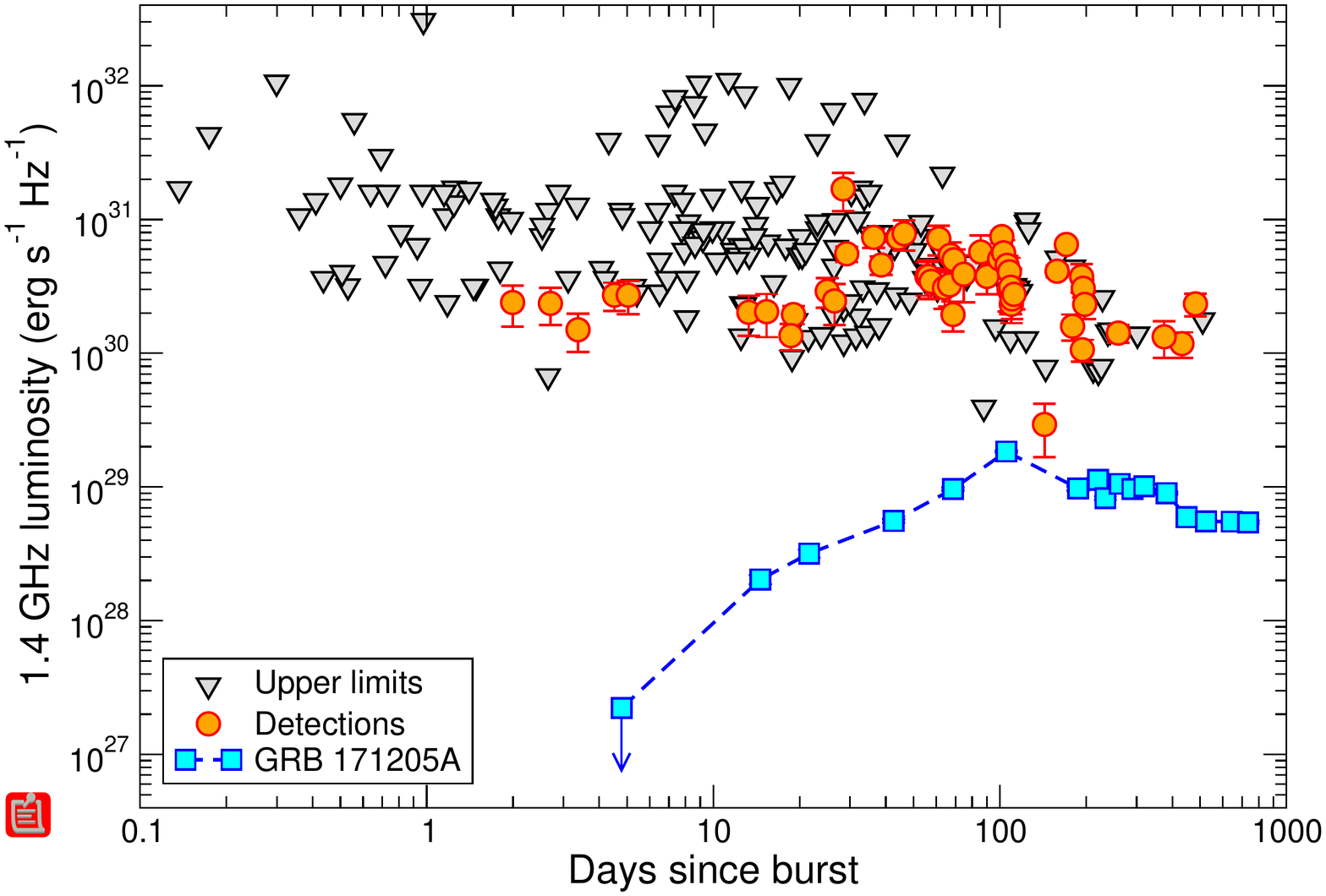}
\caption{Plot of 1.4 GHz luminosities of canonical GRBs taken from \citet{pchandra}. Here we overlay the uGMRT 1.3 GHz measurements
for GRB 171205A. Our values are at least two orders of magnitude smaller than canonical GRBs.}
\label{lband} 
\end{figure}

The peak radio flux density of GRB 171205A at 1.3 GHz is $\sim 10^{29}$ \ergs\,Hz$^{-1}$. This is two orders of  magnitude fainter than
cosmological GRBs at this frequency (Fig. \ref{lband}). However, these values are comparable to other low-luminosity GRBs, e.g. GRB 031203 \citep{soderberg+04}, 980425 \citep{kulkarni}. 

 The value of $A_*$ in the standard afterglow and the SBO models are $1.58^{+0.11}_{-0.75}$  and 
$2.89^{+1.95}_{-1.26}$, respectively, which, assuming a wind velocity of 1000\,\kms, translate to mass-loss rates of 
$1.58^{+0.11}_{-0.75}\times 10^{-6} $ M$_\odot\,\rm yr^{-1}$ and $2.89^{+1.95}_{-1.26}\times 10^{-6} $ M$_\odot\,\rm yr^{-1}$, respectively, for the two models. The nature of the surrounding ambient medium reflects on the progenitor nature  of GRBs. It is expected that the progenitors of long GRBs are massive stars (Wolf-Rayet) 
and in most of the cases a long GRB is associated with a supernova \citep{kulkarni,woos}. Another evidence for massive star progenitors is that the long GRBs generally have  star-forming host galaxies \citep{GRB}.  In such a case, one expects the association with a wind like circumburst medium. 
However, several GRBs from massive stars collapse have shown homogeneous density \citep{homo,homo1}. 
A  constant density medium can be produced around a massive star if the wind faces a shock termination \citep{ch}. 
The low frequency observations present here provide a unique opportunity to determine the nature of the circumburst medium of  GRB 171205A
and establish that GRB 171205A exploded in a wind like environment.
At uGMRT frequencies,  the  optically thick to thin transition peak arises  at a long time ($t > 100$ d) after  the burst,
indicating a relatively high density medium.  This may be created due to a large stellar mass-loss rate or a low wind velocity. Some previous works \citep{cr} have shown that the large mass-loss rate for  Wolf-Rayet stars are associated with large metallicity of the medium. Thus GRBs in wind medium can be potential tools for studying metallicity variation at different redshifts.\\

The uGMRT light curve declines as $\sim t^{-0.7}$.  This indicates that there is   no jet break until 3 years.
There are several explanations for the lack of jet breaks in some GRBs. 
In the cases of GRB 980326 and GRB 980519, \citet{gv14} have argued that a wind medium can dilute the jet break even for highly collimated bursts.
The jet break will be absent if the radio emission indeed arises from a quasi-spherical afterglow, such as that due to shock breakout \citep{ns12}
or cocoon \citep{nakar15}. In case of GRB 030329, \citet{berger+03} have argued that radio emission may be arising from two components, a narrow
jet, surrounded by a wider component (e.g. cocoon) and the radio emission is being dominated by the wider component. However, the requirement of this
model is that the contribution to the  radio afterglow from the narrow jet may be negligible. X-ray observations cover the period of $\sim 200$  days and show no indication of a  jet break at least until the last detection on day $\sim 70$. Using $t_j >71$\,d, gives a limit $\theta_j > 1.2$ radians for
the AG model and $\theta_j > 1.9$ for SBO model  \citep{nava+07,wang+18}.

\citet[][and references therein]{margutti+13} have shown that ejecta kinetic energy profiles in stripped enveloped supernovae vary based
on different explosion mechanisms. 
While
stripped-envelope supernovae have a steep dependence $E_K \propto (\Gamma \beta)^{-5.2}$ indicating
no central engine activity, relativistic supernovae, sub-energetic GRBs with SBO mechanisms are flatter with  $E_K \propto (\Gamma \beta)^{-2.4}$ showing weak 
activity from central engine. Canonical GRBs, on the other hand,  follow $E_K \propto (\Gamma \beta)^{-0.4}$  typical of jet-driven explosions with long-lasting central engines. Our radio modeling and  the relativistic treatment of \citet{bd+13}, results in
 $E_K \approx 1.1 \times 10^{51}$ ergs and $\Gamma \beta \sim1 $. 
 For the non-relativistic supernova component, we use values from \citet{izzo+19}, i.e. supernova kinetic energy
 $E_K=2.4 \times 10^{52}$ ergs and ejecta velocity $55000$ km\,s$^{-1}$.
Using these values,  $E_K \sim (\Gamma \beta)^{-1.9}$.
While GRB 171205A is a sub-energetic GRB, it follows the energy-velocity profile somewhere between canonical GRBs and
the SBOs. These arguments suggest that the jet and shock-breakout both may play
an important role in the late time afterglow emission in GRB 171205A.

Since we have radio light curve peaks at two uGMRT frequencies, we could also estimate some more 
parameter evolutions. The relativistic energy under equipartition assumption at the two epochs are $E_{\rm Eq}=3.6\times10^{48}$ ergs and $4.9\times10^{48}$ ergs  
\citep{bd+13}. Thus there is an indication
of enhancement of energy $E \propto t^{0.48}$.  This along with flatter light curve decays may also be explained if there is an energy
injection from the central engine to the shock. For an injection luminosity $L(t)=L_0(t/t_0)^{-q}$, $E \propto t^{1-q}$. This implies $q=0.52$. However,
our best fits result in much larger value of $q$ ($s=0.13$, $q=1-s\equiv 0.87$. This may imply that if energy injection, it is not continuous and probably lasted for a 
small amount of time.

The equipartition size obtained from the above formulation \citep{bd+13} at the epochs of two
peaks in band 5 and band 4  follow  R(t) $\propto$ $t^{0.49}$. We note that for ISM and Wind, density profiles, $R$ follows as $R\propto t^{1/4}$ and $R\propto t^{1/2}$, so it also points towards
a  wind like medium surrounding  GRB171205A.

Our uGMRT observations cover the period of around 1000 days. However, our data does not suggest the GRB to be in the Newtonian regime
yet (Fig. \ref{fig:radio} ).  This is not uncommon for low-luminosity GRBs  \citep[e.g. GRB 060218,][]{ic16}. 
We note that the value of $\Gamma \beta$ indicates mildly relativistic outflow. Hence it is likely that the GRB is making a transition into Newtonian regime soon.

\subsection{Shallow decay of radio afterglow}

We note that the decay of the radio afterglow is much shallower than that of the 
X-ray afterglow. 
The shallowness of radio lightcurves was first pointed out by \citet{slow}. For a reasonable
afterglow parameters, they estimated that the afterglow is supposed to cross
$\nu_m$ at around 10 days for 10 GHz and 
follow a decay slope of $(3p-1)/4$ for a wind medium, which was not the case for some GRBs, e.g. GRB 991216 and GRB 000926. 
They explored
the difference between the radio and the  optical decay indices could
be caused by the fact  that the injection frequency remains above the radio domain ($\sim$ 10\,GHz), or a different population of electrons, or a variability
of microparameters. 
Finally, they concluded that a long-lived reverse shock in the radio regime could cause this flattening. However, this is unlikely in GRB 171205A as a strong
persistent reverse shock requires a low wind density \citep{resmi}.
Here the late rising of the radio afterglow suggests a comparatively high density medium which is against the previous statement. So, this scenario can be excluded.

 \citet{kangas+19}  noted that radio afterglows of some GRBs deviate at late times and low frequencies from the standard model, 
and  attempted to explain it with the two-component jet model, a narrow jet core and a wider cocoon surrounding the jet.
\citet{granot} have explained this flattening due to counter-jet which becomes visible when turning sub-relativistic. 
While such two components should result in a bump, for stratified wind medium, the revelation of counter-jet is more gradual,
causing mild flattening. An energy injection event or a different component dominating the radio emission can also produce this flattening. 
In case of slightly off-axis jet, the early radio emission could possibly be from the cocoon, the accelerated polar ejecta and
at a late phase the contribution from the off-axis jet coming to the line of sight can increase the total radio flux \citep{decolle18}.
The lack of X-ray data at such late times
prevents  us from directly distinguishing between these scenarios.

A population of quasi-thermal
electrons has also been argued as one of the reasons \citep{warren+17}, which would mainly dominate at radio frequencies, as it would result in increased $\nu_a$ and suppressing radio emission
below this.    In GRB 171205A, this has been tentatively supported from the polarization measurements. As per \citet{urata}, the mm data revealed 0.27\% level linear-polarisation which is a factor of 4 smaller than optical polarization measurements. This has been explained as
Faraday depolarization by non-accelerated, cool electrons in the shocked region. However, one cannot rely on these results due to dispute of the detection claimed by \citet{laskar+20}.

\subsection{Origin of radio emission}

There have been suggestions that  sub-energetic bursts are simply canonical GRBs viewed
off-axis \citep{nakamura+01}. However, such bursts will have two distinguishing characters, a) low $E_p$ b) a rise in the
afterglow energy while the shocked ejecta gradually comes into our line of sight.
In GRB 171205A, the afterglow energy increases slightly  from 
$3.6\times10^{48}$\,erg to $4.9\times10^{48}$\,erg between $\sim 100$ and $\sim 200$ days. However, the $E_p$ is comparable to that of canonical GRBs.
Additionally, off-axis jet is a geometric effect, which would result in a frequency independent break in the light curve, which has not been seen for
GRB 171205A, ruling out the  off-axis model  \citep{delia+18}. Though a jet somewhat off-axis is not ruled out.
 \citet{delia+18} found that GRB 171205A   is an outlier of the Amati relation, as are some other low redshift GRBs, and its
emission mechanism should be different from that of canonical, more distant  GRBs.

There are  two models to explain the electromagnetic emission in low-luminosity GRBs, central engine driven  \citep{margutti+13,ic16} and shock-breakout driven \citep{kulkarni,ns12,bd+13,suzuki+18}. 
An issue with a purely shock breakout model is the requirement of high $\gamma$-ray efficiency, for a quasi-spherical  outflow.
Another issue is generation of such relativistic quasi-spherical outflow.
One can envisage a situation where some fraction of the supernova ejecta is accelerated to relativistic speeds to provide this 
quasi-spherical relativistic outflow. However, \citet{tan+01} have shown that only a fraction ($\sim 10^{-4}$) of supernova energy goes in
relativistic ejecta. 
\citet{nakar15} has suggested an alternative scenario where a choked jet in a low-mass envelope can  put a significant energy into a quasi-spherical, relativistic flow.

Our uGMRT model fits are incapable of distinguishing between canonical afterglow versus SBO afterglow models.
We check the applicability of this model for GRB 171205A. In this model, the expanding outflow is considered to harbour a series of successive shocks, which accelerate and catch up to the boundary and hence explain the increasing afterglow energy via continuous injection.
In this model, eventually the total energy should reach around $2.4 \times 10^{52}$ erg \citep{izzo+19}, the kinetic energy of the associated supernova. Generally one does not see such a large
 amount of energy as radio observations do not cover epochs late enough. However, our radio observations cover a period of 
 nearly 1000 days. The SBO model fit gives $E_K\approx 3.4\times 10^{50}$\, erg, two orders of magnitude smaller than 
 the one predicted in pure SBO model. 
\citet{suzuki+19}  carried out multiwavelength hydrodynamical modelling of GRB 171205A in the framework of post shock break-out relativistic SN ejecta-CBM interaction scenario.  While they claimed that this model worked well for GRB 171205A and favour the wind model, we note some problems with their model.
They had to introduce a centrally concentrated CBM with
a sudden density drop to explain the available radio and X-ray data. however, our work includes radio measurements upto 3 years and do not show a
sudden density
 drop.

The pure shock breakout model also has some other problems. A shock breakout model predicts a $\gamma$-ray emission lasting $\ge 1000 s$, lower $E_p$
(not exceeding 50 keV), 
a large absorption column density, a late time soft X-ray emission, and comparable energy in the X-ray emission and the prompt $\gamma$-ray flare.
We note that the X-ray spectrum shows an intrinsic hydrogen column density $N_H=7.4^{+4.1}_{-3.6} \times 10^{20}$\,cm$^{-2}$ \citep{delia+18}. 
This intrinsic column density is at the low-end of low-luminosity GRB distribution, even among low redshift Swift-XRT GRBs
where the mean is $N_H=2.4 \times 10^{21}$\,cm$^{-2}$ 
at $z < 0.2$ \citep{arcodia+16}. For the observations post day 1
onwards, there may be indication of a slightly higher column density $N_H=1.2^{+0.8}_{-1.7} \times 10^{21}$\,cm$^{-2}$
\footnote{\url{https://www.swift.ac.uk/xrt_live_cat/00794972/}}.
But there is no particular indication of significant
spectral softening, except of a slight indication of $\Gamma=1.63\pm0.30$ to $\Gamma=1.94\pm0.23$.
The total X-ray energy from the first observation onwards until the last detection is at least an order of magnitude smaller than the prompt energy.

\citet{delia+18} have claimed the presence of a thermal component. 
While shock  breakout from supernovae is the most favourable model  for a thermal component \citep{ns12, suzuki+18}, 
  late time photospheric emission from jet
\citep{fw13}, or thermal emission from cocoon \citep{ss13} can also explain this component.
If we assume the blackbody component to be significant, it comprises of 20\% flux and has a temperature of 89\,eV \citep{delia+18}.
This corresponds to a radius $R =(E_{\rm iso} /aT_{BB}^4)^{1/3} \approx 1.4 \times 10^{13}$ cm ($a$ is the radiation density constant). This
is much larger than the typical Wolf-Rayet star radius, but can be explained if the shock expands in a non-spherical manner. 
Alternatively, \citet{nakar15} suggested the  presence of an optically thick stellar envelope further away from the star, from where the breakout happens.

We have seen that our observations are inconsistent with a pure shock breakout due to the short duration, higher $E_p$, 
 shallow $E$ vs $\Gamma \beta$ relation, low column density and a much larger break-out radius predicted by  the
thermal component.
It can also not be explained as merely a canonical GRB seen off-axis. We show below that both these components contribute towards the radio afterglow.
GRB 171205A is the first GRB in which direct signatures of a cocoon has been seen \citep{izzo+19}. This is rare because generally  a line-of-sight jet is
much brighter than the associated  cocoon, hiding the cocoon signatures. An off-axis jet can reveal itself by the associated supernova,  but the cocoon signatures are long gone by the time supernova may be discovered. Ideally cocoon signatures are visible only  in slightly off-axis GRBs.
In GRB 171205A, the cocoon was identified by  the broad absorption features overlapped on the supernova spectrum \citep{izzo+19} 
They also estimated, from the energy deposited in cocoon,  that the jet was quite energetic \citet{izzo+19} . This may imply that we may be seeing
a slightly off-axis jet, which enabled us to reveal the cocoon, not overshadowed by the  bright jet. With this knowledge,
it is likely that radio afterglow has from both components, the sub-relativistic wider cocoon and a slightly off-axis jet.
The cocoon radio emission dominates the GRB emission at early times when the GRB jet is off-axis. Later the additional flux contribution comes when the
jet spreads sideways and comes in our line of sight.
In such a case, the total radio flux can also be large compared to on-axis GRBs since the cocoon and  the jet carry comparable energy \citep{decolle18}.

\citet{izzo+19}  could not distinguish between the cocoon along with slightly off-axis jet versus only cocoon emission,
where the faint $\gamma$-rays are the predicted signal of the cocoon breaking out of the stellar envelope.
However, our radio modelling rules out the pure shock break-out model in favour of cocoon along with  a slightly off-axis jet.
As per theoretical models, the indicated speed of the ejecta is also consistent with the sub-relativistic speeds expected in this model.

\subsection{Comparison with other low-$z$ low-luminosity GRBs}

With $E_{\rm iso}=2.3\times10^{49}$\,erg and $z=0.0368$, GRB 171205A is one of the few low-$z$ ($z\lesssim0.1$), low-luminosity GRBs. Other GRBs in this category are GRB 980425, \citep[$E_{\rm iso}=8.5\times10^{47}$ erg, $z=0.0083$;][]{galama+98},  
GRB 031203,  \citep[$E_{\rm iso}=4.0\times10^{49}$ erg, $z=0.105$;][]{sazanov+04}, 
GRB 060218 \citep[$E_{\rm iso}=2.6\times10^{49}$ erg, $z=0.033$;][]{campana+06} and
GRB 100316D \citep[$E_{\rm iso}=3.7\times10^{49}$ erg, $z=0.0590$ ;][]{starling+11}.
Like other low-luminosity GRBs, the spectrum of GRB 171205A can be fit by a simple power-law model, however, the photon   index
$\Gamma \sim 1.94$ is harder than 
these GRBs, except 
 that of GRB 031203 \citep[$\Gamma=1.63$; ][]{sazanov+04}.

Even amongst the intrinsically sub-energetic bursts, GRB 171205A resembles more closely with GRBs
980425 and 031203 and not GRBs 060218 and 100316D.  GRBs 980425 and 031203 were difficult to realise in the shock break out models due to their relatively hard spectra, shorter durations and larger $E_p$,
though lack of thermal equilibrium may accommodate it \citep{katz+10}.
GRBs 060218 and 100316D stand out due to their large durations of a few thousand seconds and lower peak energies $E_p < 50$\, keV.
The shorter $T_{90}$ and  higher $E_{\rm p}$ for GRB 171205A are 
 are comparable to the respective values for GRBs 980425 and
031203.

A major difference between GRB 171205A and other low-luminosity GRBs is the absence of intrinsic absorption  column density.
All other GRBs in this class show significant neutral hydrogen column density 
$N_H \sim (6-7)\times 10^{21}$\,cm$^{-2}$, as opposed to an order of magnitude lower $N_H$ for GRB 171205A \citep{delia+18}. The higher column density is considered an
essential feature for supernova shock break out models in low-luminosity GRBs.

The peak radio flux density of GRB 171205A at 1.3 GHz is $\sim 10^{29}$ \ergs\,Hz$^{-1}$. This value is comparable to other low-luminosity GRBs, e.g. GRB 031203 \citep{soderberg+04}. However, the X-ray luminosity at 10 hr is $L_x\approx 2\times10^{42}$ \ergs, which is 4 times smaller than that of GRB 031203 \citep[$L_x=9\times10^{42}$\,\ergs;][]{soderberg+04}.  This discrepancy is even  more significant for  a wind-like medium, since the measured peak radio luminosities are at 1.3 GHz and 8.5 GHz bands for GRB 171205A and GRB 031203, respectively.

It has been found that the X-ray afterglow decays slowly in low-luminosity GRBs. For example, 
the X-ray emission in GRB 031203 followed $F_{\rm xray}(t) \propto \nu^{-0.8} t^{-0.4}$
\citep{soderberg+04}, and for GRB 980425 $F_{\rm xray}(t) \propto \nu^{-1} t^{-0.2}$
\citep{kulkarni}. This is opposed to the canonical GRBs with  $F_{\rm xray}(t) \propto \nu^{-1.3} t^{-1}$.  \citet{bd+15} have shown that 
a flat
temporal decay of the X-ray light curve can be explained in the
shock-break out model, although dominance of an underlying supernova has also been suggested
as a possible reason for this flatness \citep{soderberg+04}. 
With  $F_{\rm xray}(t) \propto \nu^{-0.94} t^{-1.1}$, GRB 171205A is more like canonical GRBs. 
 On the contrary, the X-ray spectral index was very soft for GRB 060218A with $F_{\rm xray}(t) \propto \nu^{-2.2} t^{-1.1}$, which softened 
even further at later epochs \citep{soderberg+06}. GRB 171205A did not so that much  spectral softening. In addition, contrary to smooth light curves of
low-luminosity GRBs, the X-ray light curve of GRB 171205A revealed
three temporal breaks \citep{delia+18}.

The lack of jet break seems to be a common feature of low-luminosity GRBs. This may either indicate much wider angle ejecta responsible for afterglow emission,
or stratified medium which  can dilute the jet break in GRBs and hide its signatures.

Another defining feature of these GRBs is the presence of a thermal component, which has also been seen in
GRB 171205A \citep{delia+18}. Though this feature is quite common in  GRBs associated
with SNe \citep{campana+06, starling+11}, their origin is still a matter of debate. Shock breakout from the relativistic supernova shell is the most favourable model \citep{suzuki+18}, however,  late time photospheric emission from a jet
\citep{fw13}, or a  thermal emission from a cocoon \citep{ss13} can also explain this component.
However, it appears that the presence of  a thermal component is not a unique feature of low-luminosity GRBs. 
\citet{Starling+12, sparre+12}  analysed a sample of canonical GRBs and found the evidence of thermal component in a significant number of GRBs, though large radii associated with the blackbody emission argue against  the shock breakout model.

Finally while low-luminosity GRBs are considered to be a different class, \citet{nakar15} has provided a unified picture of low-luminosity GRBs
and  cosmological GRBs, in terms of a key difference, namely, the existence of an extended low-mass envelope. In the unified model, 
the envelope is present in the 
 low-luminosity GRBs, but absent in cosmological GRBs. 
The lack of envelope
 allows a jet to launch  without any resistance in cosmological GRBs, but the extended envelope in low-luminosity GRBs smothers the jet and 
deposits a large amount of energy in the stellar envelope, driving a mildly relativistic shock producing low-luminosity GRBs via shock breakout.
Our model does not support this picture.

\section{Summary and Conclusions}

In this paper, we have presented the sub-GHz  observations of a low-luminosity GRB 171205A upto around 1000
days. These are the best sampled low frequency  light curves of any GRB. For the first time we report a lowest frequency (250--500\,MHz) detection of a GRB. Our light curves cover a period of two years. While we are able to see the light curve peak transitions in bands 5 and 4, we missed the peak in band 3 due to the  lack of early data. The radio data suggest that the
GRB exploded in a wind medium.  At sub-GHz frequencies at late time, the afterglow is in $\nu_m<\nu_a<\nu_c$ regime, a common phenomenon seen
in late time radio afterglows. The late time {\it Chandra} X-ray measurements constrain the jet break to be $t_{\rm jet}>71$ days.

Even though  GRB\,171205A has significant similarities with other low-luminosity GRBs, it deviates from this class in many respects. We suggest that the radio emission arises from both a cocoon and a jet, where the jet is slightly off-axis \citep{izzo+19}. The early epoch radio
emission is dominated by the cocoon surrounding the jet, while the  late time radio emission has contribution from  the jet.
The flatter radio light curves, harder GRB X-ray spectrum, large $E_p$, shorter $T_{90}$, kinetic energy - ejecta velocity relation and 
time dependence of various parameters are consistent with this picture. Our work emphasises the importance of the nature of CBM, which is critical in
understanding the evolution of GRB afterglows, and are best revealed by low-frequency radio measurements.

 \acknowledgments
 
 We thank the referee for very insightful comments which helped  improve the manuscript significantly. 
 We thank Dipankar Bhattacharya for  carefully reading  the manuscript and providing critical suggestions. 
 B.M. and P.C.  acknowledge support of the Department of Atomic Energy, Government of India, under the project no. 12-R\&D-TFR-5.02-0700.
 P.C. acknowledges support from the Department of Science and Technology via SwaranaJayanti Fellowship award (file no.DST/SJF/PSA-01/2014-15). 
 We thank the staff of the GMRT that made these observations possible. GMRT is run by the National Centre for Radio Astrophysics of the Tata Institute of Fundamental Research.

\bibliography{projectGRB}{}
\bibliographystyle{aasjournal}

\end{document}